%% file: FINAL_VERSION_arXiv.tex
\documentclass[journal]{IEEEtran}
\usepackage{cite}

\usepackage{graphicx}
\usepackage{psfrag}
\usepackage{hyperref}
\usepackage{overpic}
\usepackage[tight,footnotesize]{subfigure}
\DeclareGraphicsExtensions{.xps}
\usepackage{epstopdf} 
\usepackage[cmex10]{amsmath}
\usepackage{amssymb}
\usepackage{color}
\usepackage{amsthm}

\def\tr{\mathrm{tr}}

\def\diag{\mathrm{diag}}

\newtheorem{theorem}{Theorem}
\newtheorem{lemma}{Lemma}
\newtheorem{corollary}{Corollary}

\newtheorem{assumption}{Assumption}

\usepackage[dvipsnames,svgnames]{xcolor}
\usepackage[textwidth=30mm]{todonotes}

\begin{document}

\input{main}
\input{appendices_extended}

\bibliographystyle{IEEEtran}

\bibliography{IEEEabrv,ref}

\end{document}

%% file: main.tex
%!TEX root = FINAL_VERSION_arXiv.tex
\IEEEoverridecommandlockouts
\title{{Theoretical Performance Limits of Massive MIMO with Uncorrelated Rician Fading Channels}}
\author{Luca Sanguinetti, \emph{Senior Member, IEEE}, Abla Kammoun, \emph{Member, IEEE}, and Merouane Debbah, \emph{Fellow, IEEE}
\thanks{\newline \indent L.~Sanguinetti is with the University of Pisa, Dipartimento di Ingegneria dell'Informazione, 56122, Italy (luca.sanguinetti@unipi.it). 
A. Kammoun is with the Electrical Engineering Department, King Abdullah University of Science and Technology, Thuwal, Saudi Arabia (abla.kammoun@gmail.com). M. Debbah is 
with the Mathematical and Algorithmic Sciences Lab, Huawei Technologies Co. Ltd., France (merouane.debbah@huawei.com). 
\newline \indent A preliminary version of this paper was presented at IEEE ICASSP'17, New Orleans, USA, March 2017.
\newline \indent This work was partially supported by the University of Pisa under the PRA 2018-2019 Research Project CONCEPT and also by the H2020-ERC PoC-CacheMire project (grant 727682).}\vspace{-0.5cm}}

\markboth{IEEE Transactions on Communications}%
{Shell \MakeLowercase{\textit{et al.}}: Bare Demo of IEEEtran.cls for IEEE Journals}

\maketitle

\begin{abstract}
This work considers a multicell Massive MIMO network with $L$ cells, each comprising a BS with $M$ antennas and $K$ single-antenna user equipments. Within this setting, we are interested in deriving approximations of the achievable rates in the uplink and downlink under the assumption that single-cell linear processing is used at each BS and that each intracell link forms an uncorrelated MIMO Rician fading channel matrix; that is, with a deterministic line-of-sight (LoS) path and a stochastic non-line-of-sight component describing a spatial uncorrelated multipath environment. The analysis is conducted assuming that $N$ and $K$ grow large with a given ratio $N/K$ under the assumption that the data transmission in each cell is affected by channel estimation errors, pilot contamination, an arbitrary large scale attenuation and LoS components. Numerical results are used to prove that the approximations are asymptotically tight, but accurate for systems with finite dimensions under different operating conditions. The asymptotic results are also used to evaluate the impact of LoS components. In particular, we exemplify how the number of antennas for achieving a target rate can be substantially reduced with LoS links of only a few dBs of strength. 
\end{abstract}


\section{Introduction}

Massive MIMO refers to a wireless network technology where base stations (BSs) are equipped with a very large number $M$ of antennas to serve a multitude of user equipments (UEs) by spatial multiplexing \cite{marzetta2010noncooperative,Larsson2014}. Exciting developments have occurred in the recent year. In industry, the technology has been integrated into the 5G New Radio standard \cite{Parkvall2017a}. In academia, the long-standing pilot contamination issue, which was believed to impose fundamental limitations \cite{marzetta2010noncooperative,Marzetta2016a}, has finally been resolved \cite{Bjornson2017abc}. More precisely, \cite{Bjornson2017abc} used optimal processing under correlated Rayleigh fading channel models\cite[Sect. 2.4]{massiveMIMOBook} and proved that an unbounded capacity (as $M\to \infty$) is achieved with Massive MIMO when the channel covariance matrices of the pilot contaminating UEs are asymptotically linearly independent, which is generally the case in practice.

In this work, we consider both uplink (UL) and downlink (DL) of a Massive MIMO network with $L$ cells, each comprising a BS with $M$ antennas and $K$ single-antenna UEs. We assume that the system is affected by channel estimation errors, pilot contamination, and an arbitrary large scale attenuation. Single-cell linear processing is used at the BSs \cite[Sect. 4.1.1]{massiveMIMOBook}. In particular, we assume that in the DL maximum ratio transmit (MRT) or regularized zero forcing (RZF) are used as precoding techniques, whereas maximum ratio combining (MRC) or minimum mean square error (MMSE) combing are used in the UL for data recovery. Inspired by \cite{hoydis2013massive}, we aim at deriving approximations of the achievable rates. The analysis is conducted assuming that $N$ and $K$ grow large with a non-trivial ratio $N/K$. Unlike \cite{hoydis2013massive} and most of the existing literature on the asymptotic analysis of Massive MIMO systems, we model the intracell communication links as uncorrelated Rician fading, which is more general and accurate to capture the fading variations when there is a line-of-sight (LoS) component. 

\subsection{Contributions}

Compared to Rayleigh fading, a Rician channel model makes the asymptotic analysis of Massive MIMO much more involved than in \cite{hoydis2013massive}. To overcome this issue, recent results from random matrix theory and large system analysis \cite{Hachem2012,Walid2013} are used to compute asymptotic expressions of the signal-to-interference-plus-noise ratios (SINRs), which are eventually used to approximate the achievable rates. The approximations are proven to be asymptotically tight, but accurate for realistic system dimensions, by means of the numerical results. As a notable outcome of this work, the above asymptotic analysis provides an analytical framework that can be used to evaluate the network performance under different settings without resorting to heavy Monte Carlo simulations. Also, it can be used to eventually get insights into how the LoS components in each cell affect channel estimation errors, intercell interference, and pilot contamination. To exemplify this, we consider a simplified channel model wherein the LoS vectors of different UEs are mutually orthogonal and analyze to which extent the presence of LoS components, with only a few dBs of strength, bring potential benefits to the SE of the network. To further quantify this, we numerically compute the number of antennas needed to achieve a given average rate with Rician fading and show this is substantially smaller than with Rayleigh fading. This confirms that if the UEs are scheduled properly (such that the interference between their LoS components vanish as the number of antennas grows) Rician fading may be beneficial for Massive MIMO \cite{Bjornson2016z}. 
The conference version \cite{SanguinettiGC2017} of this paper contains only a subset of the analysis above (and has no technical proofs).

\subsection{Main literature}

The main literature related to this work is represented by \cite{Wagner12,Zhang2013a,Sanguinetti_2015c, hoydis2013massive,Abla_IT_16,YueL14,Hoydis11a,Zhang13c,Zhang2014f,Said2015,Huang2015a,Zhang16,Masouros2015}. Tools from random matrix theory are used in \cite{Wagner12} to compute the ergodic sum rate in a single-cell MIMO setting with Rayleigh fading and different precoding schemes while the multicell case is analyzed in \cite{Zhang2013a}. Similar tools are used in \cite{Sanguinetti_2015c} to solve the power minimization problem under different configurations of cooperation among BSs. A similar large system analysis is presented in \cite{hoydis2013massive} for the UL and DL of Massive MIMO in cellular networks, wherein channel estimation and pilot contamination are also taken into account. All these works relies on random matrix theory tools but assume Rayleigh fading channels. With Rician fading channels, the asymptotic analysis is much more demanding with the main difficulty lying in the correlation induced by pilot contamination among the non-centered random channel estimates. In \cite{YueL14}, the authors investigate a LoS-based conjugate beamforming transmission scheme and derive some expressions of the statistical SINR under the assumption that $N$ grows large and $K$ is fixed. In \cite{Hoydis11a}, the authors study the fluctuations of the mutual information of a cooperative small cell network operating over a Rician fading channel under the form of a central limit theorem and provide an explicit expression of the asymptotic variance. In \cite{Zhang13c}, a deterministic equivalent of the ergodic sum rate and an algorithm for evaluating the capacity achieving input covariance matrices for the UL of a large-scale MIMO are proposed for spatially correlated MIMO channel with LoS components. In \cite{Zhang2014f}, the authors derive tractable
expressions for the achievable UL rate for ZF and MRC in the large-antenna
limit, along with approximating results that hold for any finite
number of antennas ($N$ grows large and $K$ is fixed). Based on these analytical results, the transmit power scaling law to meet a desirable quality of service is computed. A numerical analysis is used in \cite{Said2015} to show how LoS components may potentially improve the system performance and mitigate the pilot contamination problem. In \cite{Huang2015a}, a full-duplex multicell Massive MIMO systems is analyzed. A deterministic approximation of the UL achievable rate with MRC is derived based on random matrix theory. It is then proved that the BS-to-BS interference and self interference asymptotically vanishes. In \cite{Zhang14}, the authors study the ergodic secrecy sum rate in the DL of a multiuser MIMO system with RZF. Unlike this work where CSI is acquired by using an UL pilot training phase that induces pilot contamination, the imperfect CSI is modelled by using the generic Gauss-Markov formulation. In \cite{Zhang16}, a detailed achievable rate analysis of regular and large-scale single-user MIMO systems is presented under transceiver hardware impairments and Rician fading conditions. In \cite{Zhang2016a,Zhang2017a}, the UL SE of Massive MIMO communications with analog-digital transceivers over Rician fading channels is investigated. Unlike this work, only MRC with perfect and imperfect CSI is considered, and the asymptotic analysis is carried out in the regime where $N$ grows large and $K$ is fixed. In \cite{Bjornson2018Rician}, spatially correlated Rician fading channels are considered and the closed-form expressions for the SE with both MRC and MRT are derived with different channel estimation schemes.

\subsection{Organization and notation}
The remainder of this paper is organized as follows. Next section describes the system model and derives achievable rates for UL and DL with single-cell linear receive combining and transmit precoding. Section \ref{s-spectral-efficiency} contains our main technical results wherein we derive asymptotically tight approximations for UL and DL achievable rates. A simplified channel model is also considered to get instrumental insights into the impact of LoS components on the network performance. In Section \ref{numerical-results}, the asymptotic analysis is numerically validated. Finally, the major conclusions and implications are drawn in Section \ref{conclusions}. All the technical proofs are presented in the Appendices.

The following notation is used throughout the paper. Scalars are denoted by lower case letters whereas boldface lower (upper) case letters are used for vectors (matrices). We denote by ${\bf I}_{N}$ the identity matrix of order $N$ and call $[{\bf A}]_{i,k}$ the $(i,k)$th element of ${\bf A}$. A random vector ${\bf x} \sim \mathcal {CN}({\bf m}, {\bf C})$ is complex Gaussian distributed with mean $\bf m$ and covariance matrix $\bf C$. The trace, transpose, conjugate transpose, real part, and expectation operators are denoted by $\mathsf{tr}(\cdot)$, $(\cdot)^{T}$, $(\cdot)^{H}$,  $\mathsf{Re}\{\cdot\}$, and $\mathbb{E}\{\cdot\}$. We use $\mathbb{E}\{\cdot|\cdot\}$ to denote the conditional expectation operator. We use $a_n \asymp b_n$ to denote $a_n -b_n \to_{n\to \infty}0$ (almost surely) for two (random) sequences $a_n$, $b_n$.
\section{System Model}
Consider a Massive MIMO system composed of $L$ cells, the BS of each cell is equipped with $N$ antennas and communicates with $K$ single-antenna UEs. A double index notation is used to refer to each UE as e.g., ``user $k$ in cell $j$''. Under this convention, let ${\bf h}_{jlk} \in \mathbb{C}^{N}$ be the channel from UE $k$ in cell $l$ to BS $j$ within a coherence block.\footnote{A coherence block consists of a number of subcarriers and time samples over which the channel response can be approximated as constant and flat-fading (e.g. \cite[Sect. 2.1]{massiveMIMOBook})} We model the channel vectors $\{{\bf h}_{jjk}; k=1,\ldots, K \}$, i.e. from the $K$ UEs in cell $j$ to BS $j$, as uncorrelated Rician fading. On the other hand, an uncorrelated Rayleigh fading model is assumed for $\{{\bf h}_{jlk}; k=1,\ldots, K \}$ with $l\ne j$. This choice is motivated by the fact that the achievable rates under this model are close to those under practical measured channels with LoS and spatially distributed UEs \cite{FTufvesson15_measured_MAMIMO}. Under the above model, we have that $
{\bf h}_{jlk} = \sqrt{\beta_{jlk}}{\bf w}_{jlk}$
where $\beta_{jlk}$ accounts for the corresponding large scale channel fading or pathloss (from BS $j$ to UE $k$ in cell $l$) and ${\bf w}_{jlk} \in \mathbb {C}^{N}$ is the small scale fading channel. The channel matrix ${{\mathbf{H}}}_{jl} \in \mathbb{C}^{N\times K}$ from cell $l$ to BS $j$ is thus given by ${{\mathbf{H}}}_{jl} = [{{\mathbf{h}}}_{jl1}, \ldots, {{\mathbf{h}}}_{jlK}]$. We model ${\bf w}_{jlk}$ as
\begin{align}
{\bf w}_{jlk} = \left\{ {\begin{array}{*{20}{c}}
&\!\!\!\!{\sqrt{\frac{1}{1+\kappa_{jk}}}{\bf z}_{jjk} + \sqrt{\frac{\kappa_{jk}}{1+\kappa_{jk}}} {\bf{a}}_{jjk}}&\!\!\!\!{\quad\quad l = j}\\
&\!\!\!\!{{\bf z}_{jlk}}&\!\!\!\!{\quad\quad l \ne j}
\end{array}} \right.
\end{align}
where ${\bf z}_{jlk}\in \mathbb {C}^{N} \sim \mathcal{CN} ({\bf 0}_{N}, \mathbf{I}_N)$, ${\bf{a}}_{jjk}\in \mathbb {C}^{N}$ is a deterministic vector that accounts for the LoS component, and the scalar $\kappa_{jk} \ge 0$ is the Rician factor denoting the power ratio between ${\bf{a}}_{jjk}$ and ${\bf z}_{jjk}$. For notational convenience, we let
\begin{align}
{d}_{jlk} = \left\{ {\begin{array}{*{20}{c}}
&{\frac{\beta_{jjk}}{1+\kappa_{jk}}}&{\quad l =j}\\
&{\beta_{jlk}}&{\quad l \neq j}
\end{array}} \right.
\end{align}
such that ${\bf h}_{jlk}$ can be rewritten as
\begin{align}
{\bf h}_{jlk} = \left\{ {\begin{array}{*{20}{c}}
&{\widetilde {\bf{h}}_{jjk} + \overline {\bf{h}}_{jjk}}&{ \quad l = j}\\
&{\widetilde {\bf{h}}_{jlk} }&{\quad l \neq j}
\end{array}} \right.
%{\bf h}_{jjk} &= \widetilde {\bf{h}}_{jjk} + \overline {\bf{h}}_{jjk}\\
%{\bf h}_{jlk} &= \widetilde {\bf{h}}_{jlk} 
\end{align}
with $\widetilde {\bf{h}}_{jlk} = \sqrt{{d}_{jlk}}{\bf z}_{jlk}$ and $\overline {\bf{h}}_{jjk} = \sqrt{{d}_{jjk}\kappa_{jk}} {\bf{a}}_{jjk}$. While $\widetilde {\bf{h}}_{jjk}$ accounts for the small-scale fading variations of ${\bf h}_{jjk}$, ${\overline {\bf{h}}}_{jjk}$ depends on the large-scale fading components of propagation channel and evolves slowly in time compared to $\widetilde {\bf{h}}_{jjk}$. Measurements in \cite{Viering2002a} suggest roughly two orders of magnitude slower variations. In practice, this means that ${\overline {\bf{h}}}_{jjk}$ maintains constant for a sufficiently large number of reception phases to be accurately estimated at the BS. Therefore, in the subsequent analysis we assume that ${\overline {\bf{h}}}_{jjk}$ is
perfectly known, which is the common practice in communication theory.

\subsection{Channel estimation}
Pilot-based channel training is utilized to estimate the channel matrix ${\bf H}_{jj}$ at BS $j$. 
We assume that the BS and UEs are perfectly synchronized and operate according to a time-division duplex (TDD) protocol wherein the DL data transmission phase is preceded in the UL by a training phase for channel estimation \cite{marzetta2010noncooperative,Rusek2013, Bjornson2016z}. In Massive MIMO, it is reasonable to expect that the number of UEs per­ cell will be very large. Due to the limited number of orthogonal pilot sequences, the same set of orthogonal pilot sequences is utilized for channel estimation in each cell (i.e., the pilot reuse factor is one). This results into pilot contamination in the channel estimation \cite{marzetta2010noncooperative,Marzetta2016a,massiveMIMOBook,Bjornson2017abc}. If an MMSE estimator is used \cite{hoydis2013massive}, then the estimate ${\widehat{\mathbf{h}}}_{jlk}$ of ${\mathbf{h}}_{jlk}$ $\forall j,k$ is given by \cite{Kay1993a}
\begin{align}\label{hat_h_{jjk}}
{\widehat{\mathbf{h}}}_{jlk} =
\left\{ {\begin{array}{*{20}{c}}
&\!\!\!\!\!\!\overline {\bf{h}}_{jjk} +  \frac{{d}_{jjk}}{\frac{1}{\rho^{\rm{tr}}} + \sum\limits_{n=1}^L   {d}_{jnk}}
  \left({\bf y}_{jk}^{\rm {tr}}  - \overline {\bf{h}}_{jjk} \right)&\!\!\!\!\!\!{ \quad l = j}\\
&\!\!\!\frac{{d}_{jlk}}{\frac{1}{\rho^{\rm{tr}}} + \sum\limits_{n=1}^L   {d}_{jnk}}
  \left({\bf y}_{jk}^{\rm {tr}}  - \overline {\bf{h}}_{jjk} \right)&\!\!\!\!\!\!{\quad l \neq j}
\end{array}} \right.
\end{align}
where ${\rho^{\rm {tr}}}$ accounts for the SNR during the UL training phase and $  {\bf y}_{jk}^{\rm {tr}}$ is given by 
\begin{align}
  {\bf y}_{jk}^{\rm {tr}} = {\bf h}_{jjk}  + \sum\limits_{l=1,l\ne j}^{L}{\bf h}_{jlk} + \frac{1}{\sqrt{\rho^{\rm {tr}}}} {\bf n}_{jk}^{\rm {tr}}
\end{align}
with ${\bf n}_{jk}^{\rm {tr}} \sim \mathcal {CN}({\bf 0}_{N}, {\bf I}_{N})$. The estimate ${\widehat{\mathbf{h}}}_{jjk}$ is distributed as ${\widehat{\mathbf{h}}}_{jjk} \sim \mathcal{CN} (\overline {\mathbf{h}}_{jjk}, {\phi}_{jjk}{\bf I}_{N})  $ with 
\begin{align}
\phi_{jlk} = \frac{{d}_{jjk}{d}_{jlk}}{\frac{1}{\rho^{\rm{tr}}} + \sum\limits_{n=1}^L   {d}_{jnk}}.
\end{align}
The estimated UL channel of cell $j$ is thus given by $\widehat{{\mathbf{H}}}_{jj} = [\widehat{{\mathbf{h}}}_{jj1}, \ldots, \widehat{{\mathbf{h}}}_{jjK}]$. 
According to the orthogonality principle \cite{Kay1993a}, the estimation error ${ {\bf e}}_{jjk} = {{\bf h}}_{jjk} - {\widehat {\bf h}}_{jjk} $ is independent from ${\widehat {\bf h}}_{jjk}$ and distributed as $\sim \mathcal{CN} \left({\bf 0}_{N}, \left({d}_{jjk} - {\phi}_{jjk}\right){\bf I}_{N}\right)  $. 

\begin{figure*}\vspace{-0.8cm}
\begin{align}
\label{sinr1_{ul}}
\gamma_{jk}^{\rm{ul}} &= \frac{\vert\mathbf{v}_{jk}^{H} \widehat{\mathbf{ h}}_{jjk} \vert^{2}}{\mathbb{E}\left\{ {{\bf v}_{jk}^{H}\left(\sum\limits_{l=1, l\ne j}^{L}\sum\limits_{i=1}^{K}{\bf h}_{jli}{\bf h}_{jli}^{H} + \!\!\!\sum\limits_{i=1, i\ne k}^{K}{\bf h}_{jji}{\bf h}_{jji}^{H}+{\mathbf{ e}}_{jjk}{\mathbf{ e}}^{H}_{jjk}+\frac{1}{{\rho^{\rm{ul}}}}{\bf I}_{N}\right){\bf v}_{jk}\left| \widehat{\mathbf{ H}}_{jj} \right.} \right\}}.
\end{align} 
\hrulefill
\end{figure*}

\subsection{Uplink}

Denoting by ${\bf v}_{jk} \in \mathbb {C}^{N}$ the receiving combiner of UE $k$ in cell $j$, its received signal is given by
\begin{align}
	y_{jk}^{\rm{ul}} = \sum\limits_{i=1}^{K} {\bf v}_{jk}^{H}{\bf h}_{jji}s_{ji}^{\rm{ul}}  + \sum\limits_{l=1,l\ne j }^{L}\sum\limits_{i=1}^{K}{\bf v}_{jk}^{H}{\bf h}_{jli}s_{li}^{\rm{ul}}  + {\bf v}_{jk}^{H}{\bf n}_{jk}^{\rm{ul}}\label{y_{jk}^{ul}}
\end{align}
where $s_{li}^{\rm{ul}} \in \mathbb{C}$ is the signal transmitted in the UL from UE $i$ in cell $l$, assumed independent across $(l,i)$ pairs, of zero mean and unit variance, and ${\bf n}_{jk}^{\rm{ul}}  \sim \mathcal {CN}({\bf 0},1/{\rho^{\rm{ul}}}{\bf I}_{N})$ where $\rho^{\rm{ul}}$ accounts for the signal-to-noise ratio (SNR) in the UL. The SE that a UE can achieve is upper bounded by the channel capacity, thus an achievable SE is any number that is below the capacity. While the classical ``Shannon formula'' cannot be applied when the receiver has imperfect CSI, there are well-established capacity lower bounds that can be used. With MMSE channel estimation, a standard lower bound in Massive MIMO  \cite{hoydis2013massive,massiveMIMOBook,Marzetta2016a} allows to compute an achievable SE in the UL as $
r_{jk}^{\rm{ul}} = \mathbb{E}_{\widehat{\mathbf{ H}}_{jj}}\{\log_{2} (1+\gamma_{jk}^{\rm{ul}})\}
$
where $\mathbb{E}_{\widehat{\mathbf{ H}}_{jj}}\left\{\cdot\right\}$ denotes the expectation with respect to ${\widehat{\mathbf{ H}}_{jj}}$ and $\gamma_{jk}^{\rm{ul}}$ is the SINR given by \eqref{sinr1_{ul}} on the top of next page. As mentioned earlier, we consider MRC and single-cell MMSE (S-MMSE) as detection schemes. Then, the combiner vector ${\bf v}_{jk}$ is given by \cite{massiveMIMOBook}
\begin{align}
{\bf v}_{jk}^{\rm{MRC}}  &= \widehat{\mathbf{ h}}_{jjk} \\ 
{\bf v}_{jk}^{\rm{S-MMSE}}  &= \left(\sum\limits_{i=1}^{K}{\widehat{\mathbf{ h}}}_{jji}{\widehat{\mathbf{ h}}}^{H}_{jji} +\left( {\xi}_{j}+N\varphi_{j}^{\rm{ul}}\right){\bf I}_{N}\right)^{-1}\!\!\!\!\widehat{\mathbf{ h}}_{jjk} \label{MMSE}
\end{align}
where $\varphi_{j}^{\rm{ul}} >0$ is a design parameter and 
\begin{align}\label{xi_j}
{\xi}_{j}=\sum\limits_{l=1, l\ne j}^{L}\sum\limits_{i=1}^{K}d_{jli} + \sum\limits_{i=1}^{K}(d_{jji} - \phi_{jji}).
\end{align} 
The ``single-cell'' notion in \eqref{MMSE} is used to differentiate it from the multicell MMSE (M-MMSE) combining scheme considered in  \cite{Ngo2012b,EmilEURASIP17} for uncorrelated Rayleigh fading channels and in \cite{Bjornson2017a,Bjornson2017abc} for correlated ones. The main difference from M-MMSE is that only the channel estimates ${\widehat{\mathbf{ H}}_{jj}}$ in the own cell are computed and used in S-MMSE \cite{Bjornson2017abc,massiveMIMOBook}.
The computational complexity of S-MMSE is thus lower than with M-MMSE, though the pilot overhead is identical since the same pilots can be used to estimate both intra-cell and inter-cell channels. However, notice that M-MMSE is optimal \cite{Bjornson2017a,Bjornson2017abc} and provides unbounded capacity (in the regime $M\to \infty$ and $K$ kept fixed) for correlated Rayleigh fading channels \cite{Bjornson2017abc}. In the case of uncorrelated channels (i.e., no spatial correlation), M-MMSE combining achieves only marginal gains compared to S-MMSE and is also fundamentally limited by pilot contamination as all other 'suboptimal' single-cell processing schemes \cite{Bjornson2017a,Bjornson2017abc}. Since uncorrelated channels are considered in this work, we limit to consider S-MMSE to make the problem analytically more tractable. The asymptotic analysis of M-MMSE combining/precoding with spatially correlated Rician fading channels is interesting but left for future work since it requires more advanced random matrix theory tools.

\subsection{Downlink}
Denoting by ${\bf g}_{jk} \in \mathbb {C}^{N}$ the precoding vector of UE $k$ in cell $j$, the received signal reads 
\begin{align}
	y_{jk}^{\rm{dl}}  = \sum\limits_{i=1}^{K} {\bf h}_{jjk}^H{\bf g}_{ji}s_{ji}^{\rm{dl}}  + \sum\limits_{l=1,l\ne j }^{L}\sum\limits_{i=1}^{K}{\bf h}_{ljk}^H{\bf g}_{li}s_{li}^{\rm{dl}}  + n_{jk}^{\rm{dl}} \label{y_{jk}}
\end{align}
where $s_{li}^{\rm{ul}} \in \mathbb{C}$ is the DL data symbol intended to UE $i$ in cell $l$, assumed independent across $(l,i)$ pairs, of zero mean and unit-variance, and $n_{jk}^{\rm{dl}} \sim \mathcal {CN}(0,1/{\rho^{\rm{dl}}})$ where $\rho^{\rm{dl}}$ accounts for the signal-to-noise ratio (SNR) in the DL.  As in \cite{marzetta2010noncooperative,ngo2013energy, jose2011pilot, hoydis2013massive} (among many others), we assume that there are no downlink pilots such that the UEs do not have knowledge of the current channels but can only learn the average channel gain $\mathbb{E}\{{\bf h}_{jjk}^{H}{\bf g}_{jk}\}$. A well-established capacity lower bound that can be used within this setting is the use-and-then-forget (UatF) bound \cite[Sec. 4.3]{massiveMIMOBook}, whose name comes from the fact that channel estimates are used for designing the receive combining vectors and then effectively ``forgotten'' before signal detection. By applying the UatF bound, an achievable SE in the DL for UE $k$ in cell $j$ is obtained as  $r_{jk}^{{\rm{dl}}} = \log_{2} (1+\gamma_{jk})$
%%%%%%%%%%%%%%%%%%%%%%%%%%%%%%%%%%%
%%  
%%%%%%%%%%%%%%%%%%%%%%%%%%%%%%%%%%%
where $\gamma_{jk}^{{\rm{dl}}} $ is given by 
\begin{align}
\label{sinr1}
\gamma_{jk}^{{\rm{dl}}} = \frac{\vert \mathbb{E}\{\mathbf{h}_{jjk}^{H}\mathbf{g}_{jk}\} \vert^{2}}{\frac{1}{\rho_{\rm{dl}}}  + \sum \limits_{l=1}^{L}\sum \limits_{i=1}^{K}  \mathbb{E}\{\vert \mathbf{h}_{ljk}^{H}\mathbf{g}_{li}\vert^2\} - \vert \mathbb{E}\{\mathbf{h}_{jjk}^{H}\mathbf{g}_{jk}\} \vert^{2}}
\end{align}
where the expectation is taken with respect to the channel realizations.
The above result holds true for any precoding scheme and is obtained by treating the inter-user interference (from the same and other cells) and channel uncertainty as worst-case Gaussian noise. As said earlier, we consider MRT and RZF as precoding schemes \cite{marzetta2010noncooperative, Larsson2014, ngo2013energy, hoydis2013massive}. This yields
	\begin{align}
{\bf g}_{jk}^{\rm{MRT}}  &= \frac{\widehat{\mathbf{ h}}_{jjk}}{\sqrt{\mathbb{E}\left\{\frac{1}{K}\sum\limits_{k=1}^{K}||\widehat{\mathbf{ h}}_{jjk}||^{2}\right\}}} = {\sqrt{\theta_{j}}}\widehat{\mathbf{ h}}_{jjk}\\ 
{\bf g}_{jk}^{\rm{RZF}}  &= \frac{{\widehat{\bf u}}_{jk} }{\sqrt{\mathbb{E}\left\{\frac{1}{K}\sum\limits_{k=1}^{K}||\widehat{\mathbf{ u}}_{jk}||^{2}\right\}}}= \sqrt{\psi_{j}}{\widehat{\bf u}}_{jk}
\end{align}
where ${\widehat {\bf u}}_{jk}  = \left(\sum\limits_{i=1}^{K}{\widehat{\mathbf{ h}}}_{jji}{\widehat{\mathbf{ h}}}^{H}_{jji} +\left( {\xi}_{j}+N\varphi_{j}^{\rm{dl}}\right){\bf I}_{N}\right)^{-1}\widehat{\mathbf{ h}}_{jjk}
$
or, equivalently, ${\widehat {\bf u}}_{jk} = \frac{1}{N}{\bf Q}_{j}\widehat{\mathbf{ h}}_{jjk}$ with
\begin{align}
{\bf Q}_{j} = \left(\frac{1}{N}\sum\limits_{i=1}^{K}\widehat{\mathbf{ h}}_{jji}{\widehat{\mathbf{ h}}}^{H}_{jji} +\left(\frac{1}{N} {\xi}_{j}+\varphi_{j}^{\rm{dl}}\right){\bf I}_{N}\right)^{-1}
\end{align}
where $\varphi_{j}^{\rm{dl}}\ge 0$ is a design parameter. 

\begin{figure*}\vspace{-0.8cm}
\begin{align}
\label{sinr_{dl-MRT}}
\overline \gamma_{jk}^{\rm{MRT}}&= \frac{\overline \theta_{j} \left(\phi_{jjk} + \frac{1}{N}\overline{\bf h}_{jjk}^{H}\overline{\bf h}_{jjk}\right)^{2}}{ \underbrace{\frac{1}{N\rho^{\rm{dl}}}}_{\textnormal{Noise}}  + \underbrace{\overline s_{jk}}_{\textnormal{Non-coherent interference}} +\underbrace{\overline \theta_{j}\sum\limits_{i=1,i\ne k}^{K} \left|\frac{1}{N}\overline{\bf h}_{jji}^{H}\overline{\bf h}_{jjk}\right|^{2} + \sum\limits_{l=1,l\ne j}^{L}\overline \theta_{l} \phi_{ljk}^{2}}_{\textnormal{Coherent interference}} }\\
\label{sinr_{ul-MRC}}
\gamma_{jk}^{\rm{MRC}}  &= \frac{\left(\phi_{jjk} + \frac{1}{N}\overline {\bf{h}}_{jjk}^{H}\overline {\bf{h}}_{jjk}\right)^{2}}{ \underbrace{\frac{1}{N\rho^{\rm{ul}}\overline \theta_{j}}}_{\textnormal{Noise}}  + \underbrace{\overline s_{jk}}_{\textnormal{Non-coherent interference}}+\underbrace{\sum\limits_{i=1,i\ne k}^{K}  \left|\frac{1}{N}\overline {\bf{h}}_{jjk}^{H}\overline {\bf{h}}_{jji}\right|^{2} + \sum\limits_{l=1,l\ne j}^{L}\phi_{jlk}^{2}}_{\textnormal{Coherent interference}}}
\end{align}
\hrulefill
\end{figure*}

\section{Asymptotic spectral efficiency}\label{s-spectral-efficiency}
We exploit the statistical distribution for the channels $\{{\bf H}_{jl}\}$ and the large dimensions of $N$ and $K$ to compute a deterministic approximation of $\gamma_{jk}$ in UL and DL, which will be eventually used to find an approximation of the ergodic sum rate. In doing so, we assume the following.
\begin{assumption}\label{as:1} $N$ and $K$ grow to infinity at the same pace, that is $1 \le {\lim\inf}_N N/K\le {\lim\sup}_N N/K< \infty$.
\end{assumption}
The above assumption will be referred to as $N,K\to \infty$ in the sequel.
For technical reasons, the following reasonable assumption are is imposed \cite{hoydis2013massive, Wagner12, Hachem2012, Walid2013}.
\begin{assumption}\label{as:2}As $N\to \infty$, we have that $\forall j,l,k$ $\lim \inf_N \beta_{jlk} >0$ and $\lim \sup_N \beta_{jlk} < \infty$, and also that $\forall j$ $\lim \sup_N \;||{\bar{\bf{H}}_{jj}}|| < \infty$.
\end{assumption}
The conditions on $\beta_{jlk}$ are a well established way to model that the array gathers more energy as $N$ increases \cite{hoydis2013massive,Wagner12}. On the other hand, the condition on ${\bar{\bf{H}}_{jj}}$ implies that the Euclidean norm of the columns ${\overline {\bf{h}}}_{jjk}$ are uniformly bounded in $N,K$ \cite{Hachem2012, Walid2013}.
For simplicity, in the remainder we assume that $\varphi_{j}^{\rm{ul}} = \varphi_{j}^{\rm{dl}} = \varphi_j$ and call $\lambda_j = \frac{1}{N} {\xi}_{j}+\varphi_{j}$ with
\begin{align}\label{lambda_j}
\lambda_j  = \frac{1}{N}\sum\limits_{l=1, l\ne j}^{L}\sum\limits_{i=1}^{K}d_{jli} + \frac{1}{N}\sum\limits_{i=1}^{K}(d_{jji} - \phi_{jji}) + \varphi_{j} 
\end{align}
such that ${\bf v}_{jk}^{\rm{S-MMSE}} = {\widehat {\bf u}}_{jk} = \frac{1}{N}{\bf Q}_{j}\widehat{\mathbf{ h}}_{jjk}$ and ${\bf g}_{jk}^{\rm{RZF}} = \sqrt{\psi_{j}}{\bf v}_{jk}^{\rm{S-MMSE}}.$ Notice that \eqref{lambda_j} follows from \eqref{xi_j}.
\subsection{MRC and MRT}
Our first results are asymptotic approximations of the SINRs with MRC and MRT.
\begin{lemma}[MRT]\label{lemma_{MRT}}  Let Assumptions \ref{as:1} -- \ref{as:2} hold true. If MRT is employed, then $\gamma_{jk}^{\rm{MRT}} \asymp \overline \gamma_{jk}^{\rm{MRT}}$ with $\gamma_{jk}^{\rm{MRT}}$ given in \eqref{sinr_{dl-MRT}} where $\overline s_{jk}$ takes the form
\begin{align}\notag
\!\!\!\!\overline s_{jk}  &= \frac{1}{N}\sum\limits_{l=1}^{L}\sum\limits_{i=1}^{K} \overline \theta_{l} d_{ljk}\left({\phi}_{lli} + \frac{1}{N}\overline {\mathbf{h}}_{lli}^{H}\overline {\mathbf{h}}_{lli}\right)\\&+ \!\frac{1}{N} \!\!\sum\limits_{i=1,i\ne k}^{K} \!\!\overline \theta_{j}\left(\phi_{jji}\frac{1}{N} \overline{\bf h}_{jjk}^{H}\overline{\bf h}_{jjk}\right)\!\!
\end{align}
and $\overline \theta_{l}$ is given by 
\begin{align}\label{vartheta_{jjk}}
\overline \theta_{l} = \left(\frac{1}{K}\sum\limits_{k=1}^{K}\left({\phi}_{llk} +\frac{1}{N}\overline{\bf h}_{llk}^{H}\overline{\bf h}_{llk}\right)\right)^{-1}.
\end{align}
\end{lemma}
\begin{IEEEproof}
The proof is given in Appendix A using standard random matrix tools. Notice that this is not necessarily required for MRT since similar results can be obtained by computing the statistical expectations in \eqref{sinr1} (e.g., \cite{Bjornson2018Rician}). Unlike the former approach, the latter does not apply to MRC. This is why random matrix tools are used in Appendix A.
\end{IEEEproof}

\begin{lemma}[MRC]\label{lemma_{MRC}} Let Assumptions \ref{as:1} -- \ref{as:2} hold true. If MRC is employed, then $\gamma_{jk}^{\rm{ MRC}} \asymp \bar \gamma_{jk}^{\rm{MRC}} $ with $\gamma_{jk}^{\rm{MRC}}$ given by \eqref{sinr_{ul-MRC}} where $\overline s_{jk}$ is given by
\begin{align}\notag
\overline s_{jk} &= \frac{1}{N}\sum\limits_{l=1}^{L}\sum\limits_{i=1}^{K}  d_{jli}\left({\phi}_{jjk} + \frac{1}{N}\bar {\mathbf{h}}_{jjk}^{H}\bar {\mathbf{h}}_{jjk}\right)\\&+\frac{1}{N} \!\!\sum\limits_{i=1,i\ne k}^{K} \phi_{jjk}\frac{1}{N}\overline {\bf{h}}_{jji}^{H}\overline {\bf{h}}_{jji} 
\end{align}
and $\overline \theta_{j}$ si given by \eqref{vartheta_{jjk}}.
\end{lemma}

\begin{IEEEproof}
The proof is omitted for space limitations but relies on the same random matrix tools of those used in Appendix A for Lemma~\ref{lemma_{MRT}}. 
\end{IEEEproof}
\smallskip

As for uncorrelated Rayleigh fading channels \cite{hoydis2013massive}, the asymptotic expressions for MRT and MRC are very similar. The main difference between \eqref{sinr_{dl-MRT}} and \eqref{sinr_{ul-MRC}} is that \eqref{sinr_{ul-MRC}} is only a function of $\overline \theta_{j}$ whereas \eqref{sinr_{dl-MRT}} depends on all the factors $\{\overline \theta_{l}; l=1,\ldots,L\}$. As it follows from \eqref{vartheta_{jjk}}, the latter depends on the channel estimation quality through the coefficients $\{\phi_{jjk}\}$ and also on the normalized inner products $\frac{1}{N}\overline{\bf h}_{llk}^{H}\overline{\bf h}_{llk}$ of the LoS components in all other cells. Also, the asymptotic expressions \eqref{sinr_{dl-MRT}} and \eqref{sinr_{ul-MRC}} provide some insights into the basic behaviours of Massive MIMO systems with MRT and MRC under Rician fading channels. The signal term in the numerator of \eqref{sinr_{dl-MRT}} and \eqref{sinr_{ul-MRC}} scales quadratically with $\frac{1}{N}\overline {\bf{h}}_{jjk}^{H}\overline {\bf{h}}_{jjk}$. Following \cite{massiveMIMOBook}, the second term in the denominator is referred to as non-coherent interference because, as for Rayleigh channels, it vanishes with $1/N$ as $N\to \infty$ and $K$ is kept fixed \cite{massiveMIMOBook}. Accordingly, the third term is called coherent interference since it maintains constant with respect to $N$ as the signal term. Unlike for Rayleigh channels, it is not only a consequence of pilot contamination but also of the intracell interference generated by the LoS components. Observe that the coherent interference vanishes when pilot contamination is not present. This occurs when every UE uses a unique orthogonal pilot for channel estimation (with ensuing reduction of the system SE) or advanced pilot decontamination schemes are used \cite[Sec. 3.5]{massiveMIMOBook}.

We are ultimately interested in the ergodic achievable UL rates $r_{jk}^{\rm{dl-MRT}} = \log_{2} (1+\gamma_{jk}^{\rm{dl-MRT}})$ and $r_{jk}^{\rm{ul-MRC}}  = \mathbb{E}_{\widehat{\mathbf{ H}}_{jj}}\{\log_{2} (1+\gamma_{jk}^{\rm{ul-MRC}})\}$. Since the logarithm is a continuous function, by applying the continuous mapping theorem~\cite{Wagner12}, from the almost sure convergence results of Lemma~\ref{lemma_{MRT}} it follows that  
\begin{align}
r_{jk}^{\rm{ul-MRC}} \asymp \overline r_{jk}^{\rm{ul-MRC}} = \log_{2}\big(1 + \bar \gamma_{jk}^{\rm{MRC}} \big). 
\end{align}
Similarly, by applying the continuous mapping theorem and the dominated convergence theorem~\cite{Wagner12}, we have that  
\begin{align}
r_{jk}^{\rm{dl-MRC}} \asymp \overline r_{jk}^{\rm{dl-MRC}} = \log_{2}\big(1 + \bar \gamma_{jk}^{\rm{MRC}} \big).
\end{align}
\begin{figure*}\vspace{-0.8cm}
\begin{align}
\gamma_{jk}^{{\rm{S-MMSE}}}&= \frac{\left(1-\lambda_j[\tilde{\bf T}_j]_{kk}\right)^2}{ \underbrace{\frac{{\phi_{jjk}\overline \nu_{j} \lambda_j^2[\tilde{\bf T}_j]_{kk}^2+ \overline \varsigma_{jk}}}{N\rho^{\rm{ul}}}}_{\textnormal{Noise}} + \underbrace{\overline s_{jk}}_{\textnormal{Non-coherent interference}} + \underbrace{\lambda_j^2([\tilde{\bf T}_j^2]_{kk}-[\tilde{\bf T}_j]_{kk}^2)+\lambda_j^2\delta_j^2\sum_{l=1,l\neq j}^{L}\sum_{i=1}^{K}\phi_{jli}^2[\tilde{\bf T}_j]_{ki}[\tilde{\bf T}_j]_{ik}}_{\textnormal{Coherent interference}}
}\label{overlinegamma_{k}}\\
\label{overlinegamma_{k}_RZF}
  \gamma_{jk}^{{\rm{RZF}}}&= \frac{\overline\psi_{j}\Big(1-\lambda_j[\widetilde{\bf T}_j]_{kk}\Big)^{2}}{ \underbrace{\frac{1}{N\rho^{\rm{dl}}}}_{\textnormal{Noise}} + \underbrace{\overline s_{jk}}_{\textnormal{Non-coherent interference}} + \underbrace{\overline{\psi}_j\lambda_j^2\left([\widetilde{\bf T}_j^2]_{kk}-[\widetilde{\bf T}_j]_{kk}^2\right)+\sum\limits_{l=1,l\ne j}^{L}\overline \psi_{l}\left({\phi_{ljk}\delta_{l}}\right)^{2}\lambda_l^2[\widetilde{\bf T}_l^2]_{kk}}_{\textnormal{Coherent interference}}}
\end{align}
\hrulefill
\end{figure*}
\vspace{-0.5cm}
\subsection{S-MMSE and RZF}

To begin with, call ${{\boldsymbol{ \Phi}}}_{jj}=\diag\{\phi_{jj1},\ldots,\phi_{jjK}\}$ and rewrite $\widehat {\bf H}_{jj} = [\widehat {\bf h}_{jj1} \cdots \widehat {\bf h}_{jjK}] \in\mathbb{C}^{N\times K}$ as
\begin{align}\label{H_{28}}
\widehat {\bf H}_{jj} &= {\bf{Z}}_{jj}{{\boldsymbol{ \Phi}}}_{jj} + \overline {\bf{H}}_{jj}.
\end{align}
Then, let us introduce the fundamental equations that are needed to express an asymptotic approximation of $\gamma_{jk}$ under S-MMSE and RZF. We start with:
\begin{align}\notag
\delta_{j}&=\! \frac{1}{N} \tr \!\left( \!\lambda_{j}\!\left( 1+ \widetilde\delta_{j} \right)\!{\bf I}_N \!+\!\frac{1}{N}\overline {\bf H}_{jj} \!\left( {\bf I}_{K} \!+\! {\delta_{j}}{{\boldsymbol{ \Phi}}}_{jj} \!\right)^{-1}\!\overline {\bf H}_{jj}^H\right) ^{\!\!-1}\!\!\!\!\!\!\\&\triangleq\frac{1}{N} \tr \left({{\bf T}}_{j}\right)\label{delta_j}\\\notag
 \widetilde{\delta}_{j}  & = \frac{1}{N} \tr  \left({{\boldsymbol{ \Phi}}}_{jj}\left(\lambda_{j} \left( {\bf I}_{K}+ \delta_{j} {{\boldsymbol{ \Phi}}}_{jj}\right) \!+ \!\frac{1}{N}\frac{\overline {\bf H}_{jj}^{H}\overline {\bf H}_{jj}}{1+\widetilde \delta_{j}} \!\right)^{\!\!-1}\right)\\&\triangleq\frac{1}{N} \tr \left( {{\boldsymbol{ \Phi}}}_{jj} {\widetilde{\bf T}}_{j}\right)\label{tildedelta_j}
\end{align}
which admits a unique positive solution in the class of Stieltjes transforms of non-negative measures with support $\mathbb{R}_{+}$\cite{Hachem2012,Walid2013}. Notice that the matrices ${{\bf T}}_{j}$ and ${\widetilde{\bf T}}_{j}$\begin{align}
\!\!\!\!   {{\bf T}}_{j} &=\left( \!\lambda_{j}\!\left( 1+ \widetilde\delta_{j} \right)\!{\bf I}_N \!+\!\frac{1}{N}\overline {\bf H}_{jj} \!\left( {\bf I}_{K} \!+\! {\delta_{j}}{{\boldsymbol{ \Phi}}}_{jj} \!\right)^{-1}\!\overline {\bf H}_{jj}^H \!\right) ^{\!\!-1}\\
 \! \!\!\!   {\widetilde{\bf T}}_{j} &=  \left(\lambda_{j} \left( {\bf I}_{K}+ \delta_{j} {{\boldsymbol{ \Phi}}}_{jj}\right) + \frac{1}{N}\frac{\overline {\bf H}_{jj}^{H}\overline {\bf H}_{jj}}{1+\widetilde{\delta}_{j}} \!\right)^{-1}\hspace{-0.4cm}
\end{align}
in \eqref{delta_j} and \eqref{tildedelta_j}
are approximations of the resolvent ${\bf Q}_{jj} = (\frac{1}{N}\widehat {\bf H}_{jj}\widehat {\bf H}_{jj}^{H} + \lambda_j{\bf I}_{N} )^{-1}$ and co-resolvent ${\widetilde {\bf Q}}_{jj} = (\frac{1}{N}\widehat {\bf H}_{jj}^{H}\widehat {\bf H}_{jj}  +\lambda_j{\bf I}_{K} )^{-1}$. Also, let us define the following quantities (that will be useful in the remainder of this work):
\begin{align}\label{vartheta_{j}}
    F_{j}=(1+\tilde{\delta}_j)^{-2}\frac{1}{N^2}\tr \left(\boldsymbol{\Phi}_{jj}\widetilde{\bf T}_j\overline{\bf H}_{jj}^{H}\overline{\bf H}_{jj}\widetilde{\bf T}_j\right)
		                  \end{align}  
                  \begin{align}\label{delta_{j}}
    \Delta_{j} &= \left(1-F_{j}\right)^{2} -\lambda_{j}^{2}\frac{1}{N} \tr \left({\bf T}_{j}^{2} \right)\frac{1}{N} \tr  \left({{\boldsymbol{ \Phi}}}_{jj}{\widetilde {\bf T}}_{j}\right)^{2}
		                  \end{align}  
                  \begin{align}
\tilde{\vartheta}_j&=\frac{1}{N}\tr \big(\boldsymbol{\Phi}_{jj}\widetilde{{\bf T}}_j\boldsymbol{\Phi}_{jj}\widetilde{\bf T}_j\big)\\
\label{pippo} 
                  \overline \nu_{j} & = \frac{1}{\Delta_{j}}\frac{1}{N} \tr \left({\bf T}_{j}^{2} \right)\\
                  \notag
		\overline \varsigma_{jk}&=\frac{1-F_j}{\Delta_j} \frac{[\tilde{\bf T}_j\frac{1}{N}\overline{\bf H}_{jj}^{H}\overline{\bf H}_{jj}\tilde{\bf T}_j]_{kk}}{(1+\tilde{\delta}_j)^2} \\&+\overline \nu_j\lambda_j^2\left([\tilde{\bf T}_j\boldsymbol{\Phi}_j\tilde{\bf T}_j]_{kk}-\phi_{jjk}[\tilde{\bf T}_j]_{kk}^2\right)
		                  \end{align}                
                  \begin{align}
\xi_{jk}&=\bar{\nu}_j\lambda_j^2[\tilde{\bf T}_j\boldsymbol{\Phi}_j\tilde{\bf T}_j]_{kk} +\frac{1-F_j}{\Delta_j} \frac{[\tilde{\bf T}_j\frac{1}{N}\overline{\bf H}_{jj}^{H}\overline{\bf H}_{jj}\tilde{\bf T}_j]_{kk}}{(1+\tilde{\delta}_j)^2}\\
\zeta_{jk}&=\frac{1-F_j}{\Delta_j} \lambda_j^2 [\tilde{\bf T}_j\boldsymbol{\Phi}_j\tilde{\bf T}_j]_{kk}
+\frac{\lambda_j^2\tilde{\vartheta}_j}{\Delta_j}\frac{[\tilde{\bf T}_j\frac{1}{N}\overline{\bf H}_{jj}^{H}\overline{\bf H}_{jj}\tilde{\bf T}_j]_{kk}}{(1+\tilde{\delta}_j)^2}.
\end{align}
The following theorems represent a major result of this work.
\begin{theorem}[S-MMSE]\label{S-MMSE_Theorem}
    \label{theorem1} Let Assumptions 1 -- 2 hold true. If S-MMSE is employed, then $\gamma_{jk}^{{\rm{S-MMSE}}} \asymp \overline \gamma_{jk}^{{\rm{S-MMSE}}}$ with $\overline \gamma_{jk}^{{\rm{S-MMSE}}}$ given by \eqref{overlinegamma_{k}} where  \begin{align}  
     \overline{s}_{jk}=\left(\sum_{l=1}^{L}\frac{1}{N}\sum_{i=1}^{K}\mu_{jli} \right)\xi_{jk}+\sum_{l=1}^{L}\frac{1}{N}\sum_{i=1}^{K}\gamma_{jli}\zeta_{jk} 
 \end{align}
with
\begin{align}\label{mu} 
\mu_{jli}= \left\{ {\begin{array}{*{20}{c}}
&\!\!\!\!\!\!\!\!{d_{jji}-\phi_{jji} +\lambda_j^2[\tilde{\bf T}_j\boldsymbol{\Phi}_j\tilde{\bf T}_j]_{ii}}&\!\!\!\!\!\!\!\!{\quad l =j}\\
&\!\!\!\!\!\!\!\!{d_{jli}-\lambda_j\phi_{jli}^2\delta_j \big(2[\tilde{\bf T}_j]_{ii}+\delta_j\lambda_j[\tilde{\bf T}_j\boldsymbol{\Phi}_j\tilde{\bf T}_j]_{ii}}\big)&\!\!\!\!\!\!\!\!{\quad l \neq j}
\end{array}} \right.
\end{align}
and 
\begin{align}\label{gamma} 
\gamma_{jli}= \left\{ {\begin{array}{*{20}{c}}
&{\frac{[\tilde{\bf T}_j\frac{1}{N}\overline{\bf H}_{jj}^{H}\overline{\bf H}_{jj}\tilde{\bf T}_j]_{ii}}{(1+\tilde{\delta}_j)^2} }&{\quad l =j}\\
&{\phi_{jli}^2\delta_j^2\frac{[\tilde{\bf T}_j\frac{1}{N}\overline{\bf H}_{jj}^{H}\overline{\bf H}_{jj}\tilde{\bf T}_j]_{ii}}{(1+\tilde{\delta}_j)^2}}&{\quad l \neq j.}
\end{array}} \right.
\end{align}
 \end{theorem}
 \begin{IEEEproof}
The proof is sketched  in Appendix B and relies heavily on the techniques developed in \cite{Hachem2012} and \cite{Kammoun-18}. The presence of channel estimation errors and pilot contamination makes it necessary to apply those techniques to new random quantities whose computation in explicit form requires lengthy derivations. A large effort has been made to present the results in a simple form.
\end{IEEEproof}

\begin{theorem}[RZF]\label{RZF_Theorem}
    \label{theorem1} Let Assumptions 1 -- 2 hold true. If RZF is employed, then $\gamma_{jk}^{{\rm{RZF}}} \asymp \overline \gamma_{jk}^{{\rm{RZF}}}$ with $\overline \gamma_{jk}^{{\rm{RZF}}}$ given by \eqref{overlinegamma_{k}_RZF}
where $\overline{\psi}_{j}$ takes the form:
\begin{align}\label{overline{psi}j}
        \!\! \overline{\psi}_j = \left({\lambda_{j}^2}\overline \nu _j \frac{1}{K}\tr \boldsymbol{\Phi}_{jj}\widetilde{\bf T}_j^2+\frac{1-F_j}{\Delta_j(1+\tilde{\delta}_j)^2} \frac{1}{KN}\tr \widetilde{\bf T}_j\overline{\bf H}_{jj}^{H}\overline{\bf H}_{jj}\widetilde{\bf T}_j\right)^{-1}
\end{align}
and 
 \begin{align}  
     \overline{s}_{jk}= \sum_{l=1}^{L}\overline\psi_l\mu_{ljk}\frac{1}{N}\sum_{i=1}^{K}\xi_{il}+\sum_{l=1}^{L}\overline\psi_l\gamma_{ljk}\frac{1}{N}\sum_{i=1}^{K}\zeta_{li}.
 \end{align} 
 \end{theorem}

\begin{IEEEproof}
The proof is omitted for space limitations, but follows along the lines of Theorem \ref{S-MMSE_Theorem}.
\end{IEEEproof}

 \smallskip
Unlike the asymptotic expressions for MRC and MRT, those provided in Theorems~\ref{S-MMSE_Theorem} and~\ref{RZF_Theorem} are much more involved. The distinction between non-coherent interference and coherent terms is still doable and provides evidence of the fact that the latter depends through complicated expressions of the intracell LoS components $\{\overline{\bf h }_{jji};i=1,\ldots,K\}$. Despite being involved, when applied to practical networks such approximations are very much useful since they can be used to simulate the network behavior under different settings without to carry out extensive Monte-Carlo simulations. In fact, numerical results provided in Section \ref{numerical-results} prove that the approximations provided in Theorems~\ref{RZF_Theorem} and~\ref{S-MMSE_Theorem} are asymptotically tight, but also accurate for systems with finite dimensions. Moreover, as exemplified in the sequel and in Section~\ref{case_study}, they can be used to get important insights, with respect to CSI quality, induced interference and impact of LoS components.

\subsection{Limiting case $N \to \infty$ with $K/N \to 0$}
We now look at the limiting case in which $N \to \infty$ such that $K/N \to 0$. The following results are easily obtained from the asymptotic analysis above:
\begin{corollary}[MRC and MRT]
If $N\to \infty$ such that $K/N \to 0$, then $\overline \gamma_{jk}^{\rm{MRC}} $ reduces to:
\begin{align}
\label{sinr1_{ul-MRC}}
\overline \gamma_{jk}^{\rm{MRC}}  &= \frac{ \left(\phi_{jjk} + \frac{1}{N}\overline{\bf h}_{jjk}^{H}\overline{\bf h}_{jjk}\right)^2}{\underbrace{\sum\limits_{i=1,i\ne k}^{K} \left|\frac{1}{N}\overline{\bf h}_{jji}^{H}\overline{\bf h}_{jjk}\right|^{2} + \sum\limits_{l=1,l\ne j}^{L} \phi_{ljk}^{2}}_{\textnormal{Coherent interference}} }.
\end{align}
Also, 
$\overline \gamma_{jk}^{\rm{MRT}} $ becomes
\begin{align}
\label{sinr1_{dl-MRT}}
\overline \gamma_{jk}^{\rm{MRT}}  &= \frac{ \left(\phi_{jjk} + \frac{1}{N}\overline{\bf h}_{jjk}^{H}\overline{\bf h}_{jjk}\right)^{2}}{\underbrace{\sum\limits_{i=1,i\ne k}^{K} \left|\frac{1}{N}\overline{\bf h}_{jji}^{H}\overline{\bf h}_{jjk}\right|^{2}+ \sum\limits_{l=1,l\ne j}^{L}\frac{\overline \theta_{l}}{\overline \theta_{j}} \phi_{ljk}^{2}}_{\textnormal{Coherent interference}} }
\end{align}
with $\overline \theta_{j}$ given by \eqref{vartheta_{jjk}}. \end{corollary}
\begin{IEEEproof}
The proof follows easily from Lemmas~\ref{lemma_{MRT}} and~\ref{lemma_{MRC}}  by noticing that the non-coherent interference $\overline s_{jk} \asymp \overline \theta_j \sum\nolimits_{i=1,i\ne k}^{K} \vert\frac{1}{N}\overline{\bf h}_{jji}^{H}\overline{\bf h}_{jjk}\vert^{2}$ as $N\to \infty$ with $K/N \to 0$.
\end{IEEEproof}

\smallskip
The above corollaries show that when $N$ grows at a faster rate than $K$, differently from Rayleigh fading, the coherent interference depends also on the asymptotic behavior of the inner products $\frac{1}{N}\overline{\bf h}_{jji}^{H}\overline{\bf h}_{jjk}$. If the BS is equipped with a uniform linear array (ULA) and LoS vectors $\overline{\bf h}_{jji}$ and $\overline{\bf h}_{jjk}$ are either aligned in the complex plane or have an angular difference that scales as $1/M^{\alpha}$ with $\alpha \ge 1$, then $\frac{1}{N}\overline{\bf h}_{jji}^{H}\overline{\bf h}_{jjk}$ does not vanish asymptotically. Both cases belong to the category of scenarios for which the {\emph{favorable propagation conditions}} are not satisfied \cite{Ngo2014,Masouros2015}. Similar observations can be made under RZF and S-MMSE, as shown next.
\begin{corollary}[S-MMSE and RZF]
If $N \to \infty$ such that $K/N \to 0$, we have that $\overline \gamma_{jk}^{{\rm{S-MMSE}}}$ reduces to
\begin{align}
 \frac{\left(1-\lambda_j[\tilde{\bf T}_j]_{kk}\right)^2}{ \underbrace{\lambda_j^2([\tilde{\bf T}_j^2]_{kk}-[\tilde{\bf T}_j]_{kk}^2)+\sum_{l=1,l\neq j}^{L}\sum_{i=1}^{K}\phi_{jli}^2[\tilde{\bf T}_j]_{ki}[\tilde{\bf T}_j]_{ik}}_{\textnormal{Coherent interference}}
}
\end{align}
with 
\begin{align}\label{BB.13}
\widetilde{\bf T}_j=\left(\lambda_j {\bf I}_K+\boldsymbol{\Phi}_{jj}+\frac{1}{N}\overline{\bf H}_{jj}^{H}\overline{\bf H}_{jj}\right)^{-1}.
\end{align}
Also, $\overline \gamma_{jk}^{{\rm{RZF}}}$ reduces to
\begin{align}
   \frac{\overline\psi_{j}\Big({1-{\lambda_j[\widetilde{\bf T}_j]_{kk}}}\Big)^{2}}{{\underbrace{\overline{\psi}_j\lambda_j^2\left([\widetilde{\bf T}_j^2]_{kk}-[\widetilde{\bf T}_j]_{kk}^2\right)+\sum_{l=1,l\neq j}^{L}\overline{\psi}_l \phi_{ljk}^2[\widetilde{\bf T}_l^2]_{kk}}_{\textnormal{Coherent interference}}}  
    }
\end{align}
where $
\overline{\psi}_j=\left(\frac{1}{K}\tr \boldsymbol{\Phi}_{jj}\widetilde{\bf T}_j^2 +\frac{1}{KN}\tr \widetilde{\bf T}_j\overline{\bf H}_{jj}^{H}\overline{\bf H}_{jj}\widetilde{\bf T}_j\right)^{-1}$.
%\overline{\psi}_{j} &= \frac{1}{K}\sum\limits_{i=1}^{K}\frac{\phi_{jjk} + \frac{1}{N}\overline{\bf h}_{jjk}^{H}\overline{\bf h}_{jjk}}{\left(\lambda_{j} + \phi_{jjk} + \frac{1}{N}\overline{\bf h}_{jjk}^{H}\overline{\bf h}_{jjk}\right)^{2}}
%and also 
%  \begin{align}
%        \overline{s}_{jk}&= \sum_{l=1,l\neq j}^{L}\overline{\psi}_l \phi_{ljk}^2\left([\tilde{\bf T}_l^2]_{kk}-[\tilde{\bf T}_l]_{kk}^2\right) +\overline{\psi}_j\lambda_j^2\left([\tilde{\bf T}_j^2]_{kk}-[\tilde{\bf T}_j]_{kk}^2\right) \end{align} 
 \end{corollary}
\begin{IEEEproof}
The proof relies on observing that if $N\to \infty$ with $K/N \to 0$ then $\delta_{j} \to \lambda_{j}^{-1}$ and  $\frac{1}{N}\tr{\bf T}_{j}^2\to\lambda_j^{-2}$. Also, ${F}_{j}\to 0$, $\Delta_{j} \to  1$ and $\overline \nu_{j} \to  \lambda_{j}^{-2}$. Using these results into the expressions in Theorems~\ref{S-MMSE_Theorem} and~\ref{RZF_Theorem} completes the proof.
\end{IEEEproof}
\begin{figure*}
\begin{equation}\label{vartheta}
\vartheta=-\frac{N}{K}\frac{\kappa}{(1+\kappa)}\frac{-\big(\tilde{\delta}^{\star}\big)^2\lambda-\tilde{\delta}^\star\left(\lambda+\phi(1-\frac{3K}{N})+\frac{\kappa}{1+\kappa}+\frac{K}{N}\phi^2\frac{1+\kappa}{\kappa}\right)+\phi\frac{K}{N}-\phi^2\frac{1+\kappa}{\kappa}\frac{K}{N}}{\lambda^2\phi^3(1+\tilde{\delta}^\star)^3}
\end{equation}
\hrulefill
\end{figure*}
\subsection{Limiting case $N \to \infty$ with $K/N \to 0$ under favorable propagations}

Consider now a system in which the favorable propagation conditions are asymptotically satisfied, i.e., $\frac{1}{N}{\overline {\bf{h}}}_{jji}^{H}\overline {\bf{h}}_{jjk}\to0$ $\forall i\ne k$ as $N\to \infty$ \cite{Ngo2014,Masouros2015}. For simplicity, we only consider MRT and RZF, but similar results are obtained for MRC and S-MMSE. Then, we have that:
\begin{corollary}
If $N\to \infty$ with $K/N \to 0$ and $\frac{1}{N}{\overline {\bf{h}}}_{jji}^{H}\overline {\bf{h}}_{jjk}\to0$ $\forall i\ne k$, then:
\begin{align}
\overline \gamma_{jk}^{\rm{MRT}}  &= \frac{\overline \theta_{j} \left(\phi_{jjk} + \frac{1}{N}\overline{\bf h}_{jjk}^{H}\overline{\bf h}_{jjk}\right)^{2}}{ \sum\limits_{l=1,l\ne j}^{L}\overline \theta_{l} \phi_{ljk}^{2} } %\quad  \quad 
%\overline \gamma_{jk}^{\rm{MRC}}  = \frac{ \left(\phi_{jjk} + \frac{1}{N}\overline{\bf h}_{jjk}^{H}\overline{\bf h}_{jjk}\right)^{2}}{ \sum\limits_{l=1,l\ne j}^{L} \phi_{ljk}^{2} }
\end{align}
with $\overline \theta_{j}$ given by \eqref{vartheta_{jjk}}.
\end{corollary}
\begin{corollary}[RZF]
If $N\to \infty$ with $K/N \to 0$ and $\frac{1}{N}{\overline {\bf{h}}}_{jji}^{H}\overline {\bf{h}}_{jjk}\to0$ $\forall i\ne k$, then:
\begin{align}\notag
    \overline \gamma_{jk}^{{\rm{RZF}}}&= \frac{\overline\psi_{j}\Big(\frac{\phi_{jjk} + \frac{1}{N}\overline{\bf h}_{jjk}^{H}\overline{\bf h}_{jjk}}{\lambda_{j} + \phi_{jjk} + \frac{1}{N}\overline{\bf h}_{jjk}^{H}\overline{\bf h}_{jjk}}\Big)^{2}}{ \sum\limits_{l=1,l\ne j}^{L}\overline \psi_{l}\left(\frac{\phi_{ljk}}{\lambda_{l} + \phi_{llk} + \frac{1}{N}\overline{\bf h}_{llk}^{H}\overline{\bf h}_{llk}}\right)^{2}}\\&=\frac{\overline\psi_{j}\Big(\phi_{jjk} + \frac{1}{N}\overline{\bf h}_{jjk}^{H}\overline{\bf h}_{jjk}\Big)^{2}}{ \sum\limits_{l=1,l\ne j}^{L}\left(\frac{{\lambda_{j} + \phi_{jjk} + \frac{1}{N}\overline{\bf h}_{jjk}^{H}\overline{\bf h}_{jjk}}}{\lambda_{l} + \phi_{llk} + \frac{1}{N}\overline{\bf h}_{llk}^{H}\overline{\bf h}_{llk}}\right)^{2}\overline \psi_{l} \phi_{ljk}^{2}}
\end{align}
with $\psi_{j}$ given by
\begin{align}
    \overline{\psi}_j&=\left(\frac{1}{K}\sum\limits_{k=1}^{K}\frac{\phi_{jjk}+{\frac{1}{N}\overline{\bf h}_{jjk}^{H}\overline{\bf h}_{jjk}}}{\Big(\lambda_{j} + \phi_{jjk} + \frac{1}{N}\overline{\bf h}_{jjk}^{H}\overline{\bf h}_{jjk}\Big)^{2}}\right)^{-1}.
\end{align}
\end{corollary}

\begin{IEEEproof}
If $\frac{1}{N}\overline{\bf{h}}_{jji}^{H}\overline{\bf{h}}_{jjk}\to0$ $\forall i\ne k$, then $\widetilde{\bf T}_{j}$ in \eqref{BB.13} is diagonal with $[\widetilde{\bf T}_{j}]_{kk} = (\lambda_{j} +\phi_{jjk} + \frac{1}{N}\overline{\bf h}_{jjk}^{H}\overline{\bf h}_{jjk})^{-1}$.
\end{IEEEproof}
\smallskip
{In line with \cite{Ngo2014,Zhang2014f}, the above corollaries show that if the uncorrelated Rician fading channels result in favorable propagations, then the interference vanishes as $N$ grows unbounded for both MRT and RZF. In practice, this means that those asymptotic values can only be achieved if some UEs are dropped from service \cite{Ngo2014,Bjornson2016z}. From Corollaries 3 and 4, it also turns out that, as for Rayleigh fading channels \cite[Corollary 1]{hoydis2013massive}, the asymptotic SINRs under RZF and MRT are not necessarily the same. This is because the matrix ${\bf Q}_{j} $ with RZF depends on the correlation matrix ${{\boldsymbol{ \Phi}}}_{jj}$ through \eqref{H_{28}}.}

\section{On the effect of LoS components: A case study}\label{case_study}
To get further insights on the effect of LoS components, we consider the regime in which $N$ and $K$ grow to infinity at the same pace and assume that the channel can be simply modelled as:
\begin{align}\label{simplified_{channel}}
{\bf h}_{jlk} = \left\{ {\begin{array}{*{20}{c}}
{\sqrt{\frac{1}{1+\kappa}}{\bf z}_{jjk} + \sqrt{\frac{\kappa}{1+\kappa}} {\bf{a}}_{jjk}}&{\quad l =j}\\
{ \sqrt{\alpha}{\bf z}_{jlk}}&{\quad l \ne j}
\end{array}} \right.
%{\bf h}_{jjk} &= \sqrt{\frac{1}{1+\kappa}}{\bf z}_{jjk} + \sqrt{\frac{\kappa}{1+\kappa}} {\bf{a}}_{jjk} & \quad l =j\\ 
%{\bf h}_{jlk} & = \sqrt{\alpha}{\bf z}_{jlk}  & \quad l \ne j
\end{align}
where $\alpha \in (0,1]$ is the intercell interference factor and $\kappa$ is the Rician factor, which is assumed to be the same for all UEs in cell $j$. Using the above model, we have that
\begin{align}\label{phi_{jlk}_case_study}
\phi_{jlk} = \left\{ {\begin{array}{*{20}{c}}
{\frac{1}{(1+\kappa)^{2}}\nu}&{\quad l =j}\\
{\frac{\alpha}{1+\kappa}\nu}&{\quad l \ne j}
\end{array}} \right.
%\phi_{jjk} &= \frac{1}{(1+\kappa)^{2}}\nu& \quad l =j\\ 
%\phi_{jlk} &= \frac{\alpha}{1+\kappa}\nu& \quad l \ne j
\end{align}
with $\nu = \frac{\rho^{\rm{tr}}}{1 +\rho^{\rm{tr}} \overline L} $
%\begin{align}
%\nu = \frac{\rho^{\rm{tr}}}{1 +\rho^{\rm{tr}} \overline L} 
%\end{align}
and $\overline L = \alpha(L-1) + \frac{1}{1+\kappa}$. We assume that ${\overline {\bf H}}_{jj}$ has orthogonal columns such that the LoS matrix $\overline {\bf{H}}_{jj}$ is unitary, 
$\frac{1}{N}\overline {\bf{H}}_{jj}^{H}\overline {\bf{H}}_{jj}=\frac{\kappa}{1+\kappa}{\bf I}_{K}$. This is achieved if the vectors ${\bf{a}}_{jjk}$ are such that $\frac{1}{N}{\bf{a}}_{jjk}^{H} {\bf{a}}_{jjk}=1$ and $\frac{1}{N}{\bf{a}}_{jji}^{H}{\bf{a}}_{jjk}=0$ $\forall i\ne k$. As mentioned before, in practical networks this means that some UEs must be dropped from service \cite{Ngo2014,Bjornson2016z}.
\begin{corollary}[MRC and MRT]\label{corollary5}
{Let Assumptions 1 -- 2 hold true.} If the channel is modeled as in \eqref{simplified_{channel}} and $\frac{1}{N}\overline {\bf{H}}_{jj}^{H}\overline {\bf{H}}_{jj}=\frac{\kappa}{1+\kappa}{\bf I}_{K}$ $\forall j$, then $\overline \gamma_{jk}^{\rm{MRC}}$ and $\overline \gamma_{jk}^{\rm{MRT}}$ reduce both to
\begin{align}\label{first_approx_MRT}
&\!\!\!\!\!\!\frac{1}{\frac{1}{ \nu N \rho^{\rm{dl}}}\frac{1+\kappa}{\tau} + \frac{K}{N\nu}\Big({\overline L}\frac{1+\kappa}{\tau} + \frac{1}{\tau^{2}}\frac{\kappa}{1+\kappa}\Big) + \frac{\alpha}{\tau^{2}}\Big(\overline L -\frac{1}{1+\kappa}\Big)} \\ \label{second_approx_MRT} &\!\!\!\!\!\!=\!\!
\frac{1}{\underbrace{\frac{\overline L}{ N \rho^{\rm{dl}}}\frac{1+\kappa}{\tau}}_{\textnormal{Noise}} + \underbrace{\frac{1}{\rho^{\rm{tr}}}A}_{\textnormal{Imperfect CSI}} + \underbrace{\frac{K}{N} \overline L B}_{\textnormal{Interference}} + \underbrace{\frac{\alpha}{\tau^{2}}\Big(\overline L -\frac{1}{1+\kappa}\Big)}_{\textnormal{Pilot Contamination}}}\!\!\!
\end{align}
where $\nu = \frac{\rho^{\rm{tr}}}{1 +\rho^{\rm{tr}} \overline L} $ and $\overline L = \alpha(L-1) + \frac{1}{1+\kappa}$ and 
\begin{align}\label{tau}
\tau & = {\frac{1}{1+\kappa} + \frac{\kappa}{\nu}}\\
A &= \left(\frac{K}{N}{\overline L} + \frac{1}{N\rho^{\rm{dl}}}\right)\frac{1+\kappa}{\tau} + \frac{K}{N}\frac{1}{\tau^{2}}\frac{\kappa}{1+\kappa}\\
B &=  \overline L\frac{1+\kappa}{\tau} + \frac{1}{\tau^{2}}\frac{\kappa}{1+\kappa}.
\end{align}
\end{corollary}
\begin{IEEEproof}
Consider the asymptotic results for MRT in \eqref{sinr_{dl-MRT}}. If the channel is modeled as in \eqref{simplified_{channel}} and $\frac{1}{N}\overline {\bf{H}}_{jj}^{H}\overline {\bf{H}}_{jj}=\frac{\kappa}{1+\kappa}{\bf I}_{K}$, then $\overline \theta_{j} \big(\phi_{jjk} + \frac{1}{N}\overline{\bf h}_{jjk}^{H}\overline{\bf h}_{jjk}\big)^{2}$ reduces to $\tau\frac{\nu}{1+\kappa}$ whereas the coherent interference becomes $\frac{\alpha^2}{\tau}\frac{\nu}{1+\kappa}(L-1)$. On the other hand, the non-coherent interference reduces to $\frac{K}{N}\big( \alpha(L-1) + \frac{1}{1+\kappa} + \frac{1}{\tau}\frac{\kappa}{(1+\kappa)^2}\big) $. Putting these results together yields \eqref{first_approx_MRT} from which \eqref{second_approx_MRT} follows. Notice that if $\kappa = 0$ then $\overline \gamma_{jk}^{\rm{MRC}}$ and $\overline \gamma_{jk}^{\rm{MRT}}$ coincide with the expressions provided in \cite[Cor. 2]{hoydis2013massive}.
\end{IEEEproof}

\begin{corollary}[S-MMSE and RZF]\label{corollary6}
{Let Assumptions 1 -- 2 hold true.} Denote $\widetilde \delta^\star$ the real positive solution of the following third-order polynomial equation
\begin{align}\notag
\lambda x^3&+x^2\left(2\lambda+\phi\left(1-\frac{K}{N}\right)\right)\\&+x\left(\lambda+\phi\left(1-2\frac{K}{N}\right)+\frac{\kappa}{1+\kappa}\right)-\phi\frac{K}{N}=0\label{eq:third_polynomial}
\end{align}
with $\lambda = \frac{K}{N}\Big(\alpha (L-1) + \frac{1}{1+\kappa} - \phi\Big)$  and $\phi= \frac{\nu}{(1+\kappa)^2}$.
Define $\vartheta$ as in \eqref{vartheta} on the top of the page and \begin{equation}\label{eq:61.delta}
\Delta=\left(1-\frac{\kappa}{1+\kappa}\frac{N}{K\phi}\frac{\big(\tilde{\delta}^\star\big)^2}{(1+\tilde{\delta}^\star)^2}\right)^2-\lambda^2\vartheta \frac{N}{K}\big(\tilde{\delta}^\star\big)^2.
\end{equation}
Let $\delta^\star=\frac{\tilde{\delta}^\star}{\phi}+\frac{1-K/N}{\lambda}$. 
If the channel is modeled as in \eqref{simplified_{channel}} and $\frac{1}{N}\overline {\bf{H}}_{jj}^{H}\overline {\bf{H}}_{jj}=\frac{\kappa}{1+\kappa}{\bf I}_{K}$ $\forall j$, then $\gamma_{jk}^{\rm{RZF}}$ and $ \overline \gamma_{jk}^{\rm{S-MMSE}} $ reduce both to
\begin{align}\label{first_approx_RZF}
\frac{1}{\frac{1}{ \nu N \rho^{\rm{dl}}}\frac{1+\kappa}{X} + { \overline {s}}^{_\prime} + \frac{\alpha}{\underline{\tau}^{2}}\Big(\overline L -\frac{1}{1+\kappa}\Big)}
\end{align}
where $X^{-1} = \frac{N}{K}\left(-1+\frac{1-\frac{\kappa(1+\kappa)}{\nu}\frac{N}{K}(\tilde{\delta}^\star)^2}{\Delta}\right) \frac{1+\kappa}{\left(1-\lambda \frac{N}{K}\frac{\tilde \delta^\star}{\phi} \right)^2}
$, $\overline{s}' = \frac{\overline{s}}{\overline\psi(1-\lambda\frac{N}{K}\frac{\tilde{\delta}^\star}{\phi})^2}$ with $\frac{\overline{s}}{\overline\psi}$ given in Appendix C, and 
\begin{align}\label{theta_star}
\underline{\tau} &= \frac{1}{1+\kappa} + \frac{\kappa}{\nu}\frac{1}{\lambda\delta^\star(1+\widetilde{\delta}^\star)}.
\end{align}
\end{corollary}
\begin{IEEEproof}
The proof sketch is reported in Appendix C.
\end{IEEEproof}
\smallskip
The above results can be used to get instrumental insights. Let's consider for simplicity the asymptotic expression provided in Corollary~\ref{corollary5}.  As seen, the effective SNR, given by ${ \nu N \rho^{\rm{dl}}}\frac{\tau}{1+\kappa}$, increases linearly with $N$ as it happens for uncorrelated Rayleigh fading channels \cite[Corollary 2]{hoydis2013massive}. Also, it increases with the Rician component $\kappa$ as $\nu\frac{\tau}{1+\kappa}$; notice that $\tau$ increases as $\kappa$ grows large. As for a Rayleigh model, the channel estimation errors and interference vanish only if $N$ grows large. Indeed, if $\kappa$ increases $A$ tends to $\frac{K}{N}{\overline L} + \frac{1}{N\rho^{\rm{dl}}}$ whereas $B$ goes to $\overline L$. On the other hand, the pilot contamination term goes to zero with $\kappa$ as $1/\kappa^{2}$ (since $\tau \to \kappa/\nu$ as $\kappa$ grows large). As for uncorrelated Rayleigh fading channels, in the limiting case in which $N \to \infty$ such that $K/N \to 0$, pilot contamination remains the only performance limitation. Unlike in Rayleigh fading channels, however, it depends on the Rician factor as shown below.

\begin{corollary}
If $N\to \infty$ with $K/N \to 0$, then $\overline \gamma_{jk}^{\rm{MRC}} = \overline \gamma_{jk}^{\rm{MRT}} = \overline \gamma_{jk}^{\rm{S-MMSE}} = \overline \gamma_{jk}^{\rm{RZF}} \asymp  \gamma_{\infty} $ with
\begin{align}
 \gamma_{\infty}= \frac{1}{\frac{\alpha}{\tau^{2}}\Big(\overline L -\frac{1}{1+\kappa}\Big)} = \frac{\Big({\frac{1}{1+\kappa} + \frac{\kappa}{\nu}}\Big)^2}{\alpha^2(L-1)} 
\end{align}
and the ultimately achievable rate is given by
\begin{align}
R_{\infty} = \log_2\Big(1 + \gamma_{\infty}\Big)  = \log_2\left(1 + \frac{\Big({\frac{1}{1+\kappa} + \frac{\kappa}{\nu}}\Big)^2}{\alpha^2(L-1)} \right).
\end{align}
\end{corollary}
\begin{IEEEproof}
It follows from Corollaries~\ref{corollary5} and~\ref{corollary6} by taking the limit $N\to \infty$ with $K/N \to 0$. {Specifically, with S-MMSE and RZF, if $K/N \to 0$ then $\widetilde{\delta}^\star \to 0$ and ${\delta}^\star =\frac{\tilde{\delta}^\star}{\phi}+\frac{1-K/N}{\lambda} \to \lambda$. Also, we have that $X \to 1$ and $\bar s^\prime \to 0$. Moreover, $\underline{\tau}$ in \eqref{theta_star} tends to $\frac{1}{1+\kappa} + \frac{\kappa}{\nu}$. This completes the proof.}
\end{IEEEproof}
\smallskip
Notice that if $\kappa = 0$ then $R_{\infty}  = \log_2\big(1 + \frac{1}{{\alpha}^2( L -1)}\big)$ coincides with the ultimately achievable rate provided in \cite[Eq. (32)]{hoydis2013massive}. On the other hand, if $\kappa \gg 1$ we have that ${\frac{1}{1+\kappa} + \frac{\kappa}{\nu}} \approx \frac{1 +\rho^{\rm{tr}} \alpha(L-1)}{\rho^{\rm{tr}}}{\kappa}$ since $\nu \approx \frac{\rho^{\rm{tr}}}{1 +\rho^{\rm{tr}} \alpha(L-1)} $ and thus
\begin{align}
R_{\infty}   \approx  \log_2\left(1 + \Big(\frac{1}{\rho^{\rm{tr}}\alpha}+L-1\Big)^2\frac{\kappa^2}{L-1}\right)
\end{align}
scales logarithmically with $\kappa$ as $2\log_2(\kappa)$.

\begin{figure}[t]
     \begin{center}
         \psfrag{Number of antennas (N)}[c][b]{\footnotesize{Number of antennas, $N$}}
    \psfrag{Average sum SE [bit/s/Hz/UE]}[c][t]{\footnotesize{Spectral efficiency [bit/s/Hz/UE]}}
     \psfrag{MRC Approx}[l][l]{\scriptsize{MRC Approx}}
     \psfrag{S-MMSE Approx}[l][l]{\scriptsize{S-MMSE Approx}}
     \psfrag{MRC Sim}[l][l]{\scriptsize{MRC Sim}}
     \psfrag{S-MMSE Sim}[l][l]{\scriptsize{S-MMSE Sim}}
      \psfrag{kappa 1}[l][l]{\hspace{-0.1cm}\footnotesize{\quad$\kappa = 4$}}
      \psfrag{kappa 2}[l][l]{\hspace{-0.1cm}\footnotesize{\quad$\kappa = 0$}}
            \psfrag{Infinite rate1}[l][l]{\footnotesize{\quad$R_{\infty}$}}
                  \psfrag{Infinite rate2}[l][l]{\footnotesize{\quad$R_{\infty}$}}
       {
           
            \includegraphics[width=9.cm]{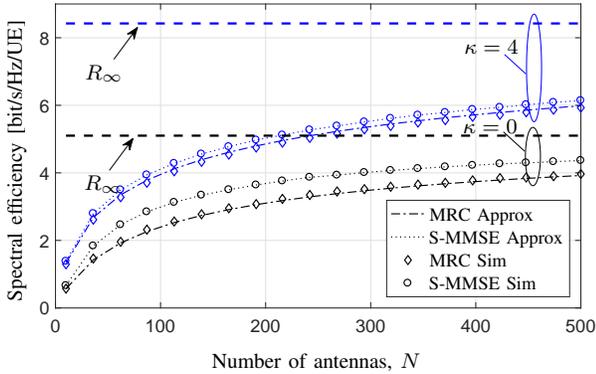}
        }
    \end{center}
    \caption{SE per UE with MRC and S-MMSE as a function of the number of antennas $N$ for the simplified channel given in \eqref{simplified_{channel}} with $L=4$, $K=10$, $\alpha =0.1$, $\rho^{\rm {tr}} = 6$ dB,  $\rho = 10$ dB, and $\kappa =0$ or $4$.}%
  \label{figure1}
\end{figure}

Before proceeding further, let's us validate the accuracy of the approximations provided in Corollaries~\ref{corollary5} and~\ref{corollary6}. To this end, we assume that the antenna array is uniform and linear with half-wavelength antenna spacing and model ${\bf{a}}_{jjk}$ as $
{\bf{a}}_{jjk} = \big[1, e^{-i\pi\sin(\vartheta_{jjk})},\ldots,e^{-i\pi (N-1)\sin(\vartheta_{jjk})}\big]^T$
where $\vartheta_{jjk}$ denotes the azimuth angle to UE $k$ in any cell $j$. Following~\cite{Ngo2014}, we assume that the each BS can create $N$ orthogonal beams with angles $\{\vartheta_{jjk}\}$ such $\sin(\vartheta_{jjk}) = -1 + \frac{2k-1}{N}$ and assume that each one of the $K$ UEs is randomly and independently assigned to one of them. 

 \begin{figure}
 \centering
   {         \psfrag{Rician factor}[c][b]{\scriptsize{Rician factor, $\kappa$}}
    \psfrag{Number of antennas}[c][t]{\scriptsize{Number of antennas, $N$}}
     \psfrag{rate1rate1rate1}[l][l]{\scriptsize{\!$\overline R \!= \!1$ bit/s/Hz}}
     \psfrag{rate2rate2}[l][l]{\scriptsize{\!$\overline R \!= \!1.5$ \!bit/s/Hz}}
          \psfrag{rate3rate3}[l][l]{\scriptsize{\!$\overline R \!= \!2$ bit/s/Hz}}\includegraphics[width=9cm]{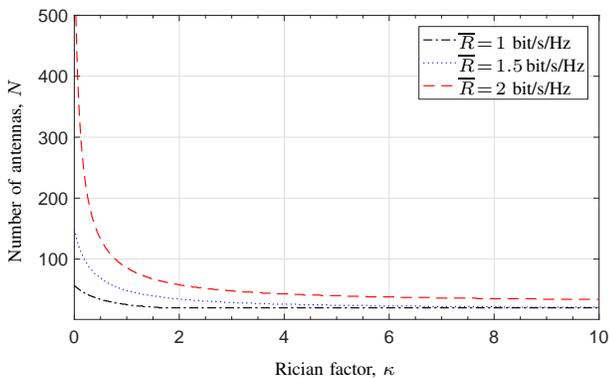}}
 \caption{Number of antennas needed with MRC and the simplified channel given in \eqref{simplified_{channel}} to achieve SE per UE of $\overline R$ bit/s/Hz/UE versus the Rician factor $\kappa$ with $L=4$, $K=10$, $\rho^{\rm {tr}} = 6$ dB,  $\rho = 10$ dB and $\alpha =0.3$.}\label{figure2}
 \end{figure}

Fig.~\ref{figure1} reports the average rate per UE of MRC and S-MMSE as a function of $N$ for $L=4$, $K=10$, $\alpha =0.1$, $\rho^{\rm {tr}} = 6$ dB,  $\rho = 10$ dB, and $\kappa =0$ or $4$. Markers are obtained
using Monte-Carlo (MC) simulations whereas the lines are obtained using the closed-form approximations of Corollaries~\ref{corollary5} and~\ref{corollary6}.  As seen, the approximations match perfectly with the MC simulations for any $N$. This proves that the asymptotic approximations are not only asymptotically tight, but 
accurate even for networks of finite size. Both schemes provide higher rates when $\kappa =4$. As expected, S-MMSE outperforms MRC. This is achieved at the price of a higher computational complexity. With $\kappa =4$, the gain of S-MMSE is only $3-6\%$. This means that LoS components may allow not only to achieve higher rates but also to use schemes with lower complexity. As predicted by the analytical results, $R_\infty$ increases as $\kappa$ grows. With $\kappa =4$, $R_\infty$ is increased by a factor $1.65$ compared to the Rayleigh fading case (i.e., $\kappa =0$). However, a larger number of antennas is needed to approach $R_\infty$ when $\kappa$ increases. With $N=500$, the S-MMSE achieves $85\%$ and $70\%$ of $R_\infty$ with  $\kappa =0$ and $4$, respectively.

  \begin{figure}
 \centering
   {         \psfrag{Intercell interference}[c][b]{\footnotesize{Intercell interference, $\alpha$}}
    \psfrag{Number of antennas}[c][t]{\footnotesize{Number of antennas, $N$}}
     \psfrag{rate1rate1}[l][l]{\scriptsize{$\kappa = 0$ }}
     \psfrag{rate2rate2}[l][l]{\scriptsize{$\kappa = 1/2$}}
          \psfrag{rate3rate3}[l][l]{\scriptsize{$\kappa = 4$}}
          \psfrag{rate4rate4}[l][l]{\scriptsize{$\kappa = 10$}}
          \includegraphics[width=9cm]{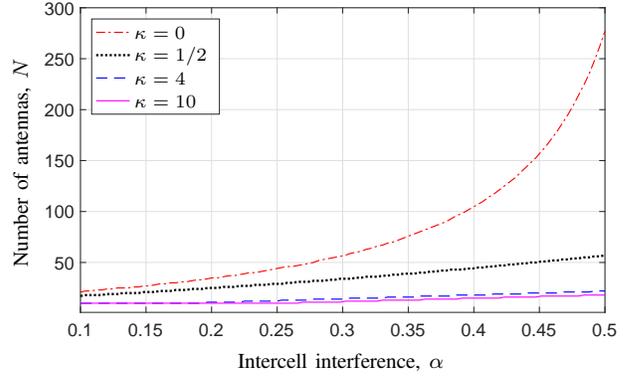}}
 \caption{Number of antennas needed with MRC and the simplified channel given in \eqref{simplified_{channel}} to achieve a SE per UE of $\overline R=1$ bit/s/Hz/UE versus the intercell interference factor $\alpha$ with $L=4$, $K=10$, $\rho^{\rm {tr}} = 6$ dB,  $\rho = 10$ dB.}\label{figure3}
 \end{figure}

Fig.~\ref{figure2} shows the number of antennas $N$ that is needed with MRC to achieve a given spectral efficiency of $\overline R$ bit/s/Hz per UE. We consider $L=4$, $K=10$, $\rho^{\rm {tr}} = 6$ dB, $\rho = 10$ dB and an intercell interference factor $\alpha$ of $0.1$ or $0.3$. The curves are obtained using the closed-form approximation of Corollary~\ref{corollary5} and show the impact of the LoS components in reducing $N$. Compared to the Rayleigh fading case (i.e., $\kappa =0$), when $\overline R=2$ bit/s/Hz, $\kappa=4$ and $\alpha=0.3$, $N$ can be roughly reduced by a factor of $10$. Fig.~\ref{figure3} illustrates the impact of the LoS components when the intercell interference increases in the same setting of Fig.~\ref{figure2} for $\bar R=1$ bit/s/Hz. Compared to the case with $\kappa=0$ where an exponential increase of $N$ is observed as $\alpha$ grows, a relatively slow increase is observed in the presence of LoS components. A Rician coefficient of $\kappa =1/2$ is enough to reduce the number of antennas of a factor ranging from $1.25$ to $4.5$.

\begin{figure}[t!]
\begin{center}
\begin{overpic}[width=9cm]{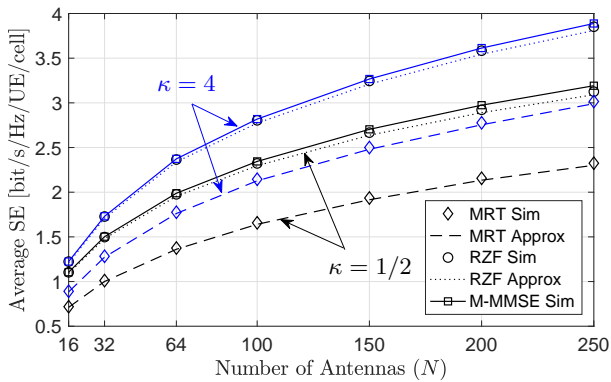}
\put(26,43){{\small{\color{blue}$\kappa = 4$}}}
\put(50,16){ {\small$\kappa = 1/2$}}
\end{overpic}
\end{center}
\caption{SE per UE with MRT and RZF versus $N$ when $L=4$, $K=10$, and the Rician factor is the same for all UEs and equal to $\kappa =1/2$ and $4$.} \label{figure5}
\end{figure}

\section{Numerical validation of the asymptotic analysis}\label{numerical-results}
MC simulations are now used to validate the accuracy of the above asymptotic analysis of Lemma~\ref{lemma_{MRT}} and Theorem~\ref{RZF_Theorem} for finite values of $N$ and $K$. The Matlab code available online at \url{https://github.com/lucasanguinetti/} enables further testing. Similar trends are obtained for the UL but are omitted for space limitations. We consider a multicell system with $L=4$ cells, with each covering a square area of $250\times 250$ m. A wrap around topology is used to simulate that all BSs receive equally much interference from all directions. The parameter $\beta_{jlk}$ is modeled in decibels as $\beta_{jlk}^{j} = \Upsilon - 10 \alpha \, \log_{10} \left( \frac{x_{jlk}}{1\,\textrm{km}} \right) + \Psi_{jlk}$
where $x_{jlk}$\,[km] is the distance between the transmitter and the receiver, the \emph{pathloss exponent} $\alpha=3.7$ determines how fast the signal power decays with the distance, and $\Upsilon = -148$ dB determines the median channel gain at a reference distance of 1\,km. Also, $\Psi_{jlk} \sim \mathcal{N}(0,\sigma_{\mathrm{sf}}^2)$ with $\sigma_{\mathrm{sf}} = 10$ accounts for the shadow fading. We assume that $K=10$ UEs are randomly and
uniformly distributed in each cell, at distances larger than 35\,m from the BS. Results are averaged over 50 UE distributions. We consider communication over a 20\,MHz bandwidth with a total receiver noise power \index{noise} of $-94$\,dBm. The median SNR of a UE at 35\,m from its serving BS is $20.6$\,dB, while a UE in any of the corners of a square cell gets $-5.8$\,dB. We consider a uniform linear array with half-wavelength antenna spacing for which $
{\bf{a}}_{jjk} = \big[1, e^{-i\pi\sin(\vartheta_{jjk})},\ldots,e^{-i\pi (N-1)\sin(\vartheta_{jjk})}\big]^T$. The angles $\{\vartheta_{jjk}\}$ are randomly and independently chosen in the interval $[0,2\pi]$. For simplicity, the Rician factor is the same for all UEs, i.e., $\kappa_{jk} = \kappa$ $\forall j,k$.

\begin{figure}[t!]
\begin{center}
\begin{overpic}[width=9cm]{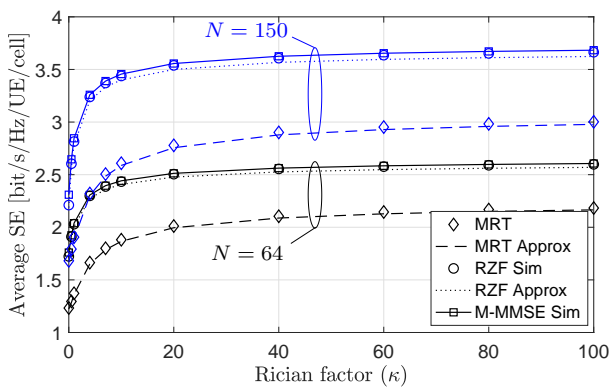}
\put(33,52){\footnotesize{{\color{blue}$N=150$}}}
\put(34,19){\footnotesize{$N = 64$}}
\end{overpic}
\end{center}
\caption{SE per UE with MRT and RZF vs $\kappa$ when $L=4$, $K=10$ and $N=64$ or $150$.} \label{figure5}
\end{figure}

Fig.~\ref{figure5} illustrates the average SE per UE when $N$ grows large with MRT and RZF. For completeness, comparisons are made with M-MMSE precoding \cite{massiveMIMOBook,Bjornson2017abc}. The Rician factor for all UEs is $\kappa=1/2$ and $4$; that is, the LoS vectors are respectively $-3$ dB and $6$ dB stronger than Rayleigh vectors. As seen, the asymptotic results perfectly match the MC simulations. RZF provides higher SE than MRT and achieves the same performance with M-MMSE (for the considered setup) for any value of $N$ and $\kappa$. As $\kappa$ increases, both RZF and MRT provide better performance. This is further investigated for $N=64$ or $150$ in Fig.~\ref{figure5}. The performance gap between RZF and MRT reduces in both cases quite rapidly as $\kappa\le20$; this is because, on average, the intra-cell interference reduces when $\kappa$ increases. However, the gap reduces much more slowly for $40 \le \kappa\le100$.

\section{Conclusions}\label{conclusions}
We investigated the effect of uncorrelated Rician fading channels on the UL and DL ergodic achievable rates of MRT/MRC and RZF/S-MMSE in Massive MIMO under the assumption of channel estimation errors and pilot contamination. Recent results from random matrix theory were used to find asymptotic approximations for S-MMSE/RZF that depend only on the long-term channel statistics, the Rician factors and the deterministic components. Numerical results indicated that these approximations are asymptotically tight, but also accurate for systems with finite dimensions. Applied to practical networks, such results can be used to simulate the network behavior without to carry out extensive Monte-Carlo simulations and get important insights into the system behavior,  with respect to the LoS vectors, CSI quality and induced interference. For a simplified channel model with orthogonal LoS components across UEs, we analytically evaluated the impact of the Rician factor of each UE on both the residual interference, induced by channel estimations errors, and pilot contamination. Also, we determined numerically how the number of antennas can be reduced for achieving a given target rate.

To make the problem analytically more tractable, single-cell processing and uncorrelated Rician fading channels were only considered in this work. Spurred by the new results in \cite{Bjornson2017abc} (that disproved previous belief on the fundamental limits of Massive MIMO), an important follow-up of this work is to consider optimal M-MMSE combining/precoding and spatially correlated Rician fading channels. This latter case can potentially be addressed by extending the random matrix theory tools developed in \cite{Kammoun-18} but applied to the DL of a single-cell multiuser large-scale MIMO network under spatially correlated Rician fading, with a common spatial channel correlation matrix among UEs.

%% file: appendices_extended.tex
%!TEX root = FINAL_VERSION_arXiv.tex
\section*{Appendix A \\ Asymptotic Analysis of MRT}
We start dividing the numerator and denominator of $\gamma_{jk}^{{\rm{dl}}}$ by $1/N$. Then, we obtain
\begin{align}
\frac{ {{\theta_{j}}{N}}\vert \mathbb{E}\{\frac{1}{N}\mathbf{h}_{jjk}^{H}{\widehat {\mathbf{h}}}_{jjk}\} \vert^{2}}{\frac{1}{N \rho_{\rm{dl}}}  + \sum \limits_{l=1}^{L}\sum \limits_{i=1}^{K}  {\theta_{l}}{N}\mathbb{E}\{\vert \frac{1}{N}\mathbf{h}_{ljk}^{H}\mathbf{\widehat h}_{lli}\vert^2\} - {\theta_{j}}{N}\vert \mathbb{E}\{\frac{1}{N}\mathbf{h}_{jjk}^{H}{\widehat {\mathbf{h}}}_{jjk}\} \vert^{2}}. 
\end{align}

\emph{1) Signal Power}: Using straightforward computations yields
$
\frac{1}{N}\mathbf{h}_{jjk}^{H} \widehat{\mathbf{h}}_{jjk}\asymp  {\phi}_{jjk} +\frac{1}{N}\overline {\bf{h}}_{jjk}^{H}\overline {\bf{h}}_{jjk}
$
and
\begin{align}
{\theta_{j}}{N}  \asymp \left(\frac{1}{K}\sum\limits_{k=1}^{K} \left({\phi}_{jjk} +\frac{1}{N}\overline {\bf{h}}_{jjk}^{H}\overline {\bf{h}}_{jjk}\right)\right)^{-1}.
\end{align}
Therefore, we have that
\begin{align}
 {{\theta_{j}}{N}}\vert \mathbb{E}\{\frac{1}{N}\mathbf{h}_{jjk}^{H}{\widehat {\mathbf{h}}}_{jjk}\}\vert^{2} \asymp \overline \theta_{j} \left({\phi}_{jjk} +\frac{1}{N}\overline {\bf{h}}_{jjk}^{H}\overline {\bf{h}}_{jjk}\right)^{2}
 \end{align}
 with $\overline \theta_{j} = \big(\frac{1}{K}\sum\nolimits_{k=1}^{K} ({\phi}_{jjk} +\frac{1}{N}\overline {\bf{h}}_{jjk}^{H}\overline {\bf{h}}_{jjk})\big)^{-1}.$
 
\smallskip
\emph{2) Interference Power}: We proceed computing the deterministic equivalent of the interference power. To begin with, we rewrite it as follows:
\begin{align}\notag
&\sum \limits_{l=1}^{L}\sum \limits_{i=1}^{K}  {\theta_{l}}{N}\mathbb{E}\{\vert \frac{1}{N}\mathbf{h}_{ljk}^{H}\mathbf{\widehat h}_{lli}\vert^2\} - {\theta_{j}}{N}\vert \mathbb{E}\{ \frac{1}{N}\mathbf{h}_{jjk}^{H}{\widehat {\mathbf{h}}}_{jjk}\} \vert^{2} \\&=  s_{jk}^{(I)} +  s_{jk}^{(P)} + {\theta_{j}}{N}\mathrm{var}\left[\frac{1}{N}\mathbf{h}_{jjk}^{H}{\widehat {\mathbf{h}}}_{jjk}\right]
 \end{align}
 where $ s_{jk}^{(I)} $ accounts for the intracell and intercell interference:
      \begin{align}
 s_{jk}^{(I)} &= \sum \limits_{l=1}^{L}{\theta_{l}}{N}\sum \limits_{i=1,i\ne k}^{K} \mathbb{E}\{\vert \frac{1}{N}\mathbf{h}_{ljk}^{H}\mathbf{\widehat h}_{lli}\vert^2\}
  \end{align}
 and $  s_{jk}^{(P)} $   is due to pilot contamination:
        \begin{align}
  s_{jk}^{(P)} &=\sum \limits_{l=1,l\ne j}^{L} {\theta_{l}}{N}\mathbb{E}\left\{\vert \frac{1}{N}\mathbf{h}_{ljk}^{H}\mathbf{\widehat h}_{llk}\vert^2\right\}.
  \end{align}
We start computing an asymptotic expression for $s_{jk}^{(I)}$. To this end, we write $s_{jk}^{(I)}$ as follows:
      \begin{align}\notag
 s_{jk}^{(I)} &= \sum \limits_{l=1,l\ne j}^{L}{\theta_{l}}{N}\sum \limits_{i=1,i\ne k}^{K} \mathbb{E}\{\vert \frac{1}{N}\mathbf{h}_{ljk}^{H}\mathbf{\widehat h}_{lli}\vert^2\} \\&+ {\theta_{j}}{N}\sum \limits_{i=1,i\ne k}^{K} \mathbb{E}\{\vert \frac{1}{N}\mathbf{h}_{jjk}^{H}\mathbf{\widehat h}_{jji}\vert^2\}.\label{A.6732}
  \end{align}
Consider the first term in \eqref{A.6732}. Since $\widehat {\mathbf{h}}_{lli}$ in \eqref{hat_h_{jjk}} is independent from $\mathbf{h}_{ljk}$ and $\mathbb{E}[\mathbf{h}_{ljk}^{H}\mathbf{h}_{ljk}]= d_{ljk}$, it easily follows that:
\begin{align}\notag
\frac{1}{N^{2}}\vert \mathbf{h}_{ljk}^{H}\widehat {\mathbf{h}}_{lli}\vert^2  &\asymp\frac{1}{N^{2}}\tr\left( d_{ljk}\left({\phi}_{lli}{\bf I}_{N} + \overline {\mathbf{h}}_{lli}\overline {\mathbf{h}}_{lli}^{H}\right)\right) \\&\asymp\frac{1}{N}d_{ljk}\left( {\phi}_{lli} + \frac{1}{N}\overline {\mathbf{h}}_{lli}\overline {\mathbf{h}}_{lli}^{H}\right).
\end{align}
Therefore, we have that:
\begin{align}\notag
 &\sum \limits_{l=1,l\ne j}^{L}{\theta_{l}}{N}\sum \limits_{i=1,i\ne k}^{K} \mathbb{E}\{\vert \frac{1}{N}\mathbf{h}_{ljk}^{H}\mathbf{\widehat h}_{lli}\vert^2\}  \\&\asymp\frac{1}{N}\sum \limits_{l=1,l\ne j}^{L} \sum\limits_{i=1, i\ne k}^{K}\overline \theta_{l}d_{ljk}\left( {\phi}_{lli} + \frac{1}{N}\overline {\mathbf{h}}_{lli}\overline {\mathbf{h}}_{lli}^{H}\right).\label{A.20}
\end{align}
Consider the second term in \eqref{A.6732} and observe that
\begin{align}\notag
\!\!\!\!\!&\vert \frac{1}{N}\mathbf{h}_{jjk}^{H}\mathbf{\widehat h}_{jji}\vert^2\\&\asymp \frac{1}{N}\left(d_{jjk}\frac{1}{N}\widehat{\mathbf{h}}_{jji}^{H}\mathbf{\widehat h}_{jji} + \frac{1}{N}\overline{\mathbf{h}}_{jjk}^{H}\widehat{\mathbf{h}}_{jji}\mathbf{\widehat h}_{jji}^{H}\overline{\mathbf{h}}_{jjk}\right) \\\notag &\asymp \frac{1}{N}d_{jjk}\left( {\phi}_{jji} + \frac{1}{N}\overline {\mathbf{h}}_{jji}\overline {\mathbf{h}}_{jji}^{H}\right) \\&\quad\quad+ \frac{1}{N}\left(\frac{1}{N}\overline{\mathbf{h}}_{jjk}^{H}\left({\phi}_{jji}{\bf I}_{N} + \overline {\mathbf{h}}_{jji}\overline {\mathbf{h}}_{jji}^{H}\right)\overline{\mathbf{h}}_{jjk}\right) \\\notag&
\asymp \frac{1}{N}d_{jjk}\left( {\phi}_{jji} + \frac{1}{N}\overline {\mathbf{h}}_{jji}\overline {\mathbf{h}}_{jji}^{H}\right) \\&\quad\quad+ \frac{1}{N}\left(\frac{1}{N}{\phi}_{jji}\overline{\mathbf{h}}_{jjk}^{H}\overline{\mathbf{h}}_{jjk}\right) + \frac{1}{N}\left(\frac{1}{N}\vert\overline {\mathbf{h}}_{jji}^{H}\overline{\mathbf{h}}_{jjk}\vert^{2}\right).\!\!
\end{align}
\begin{figure*}
\begin{align}\label{gamma_jk_App1}
\gamma_{jk}^{\rm{ul}} &= \frac{\vert\frac{1}{N}\widehat{\mathbf{h}}_{jjk}^{H}{\bf Q}_{j} \widehat{\mathbf{ h}}_{jjk} \vert^{2}}{\mathbb{E}\left\{ {\sum\limits_{l=1, l\ne j}^{L}\sum\limits_{i=1}^{K}\vert\frac{1}{N}\widehat{\mathbf{h}}_{jjk}^{H}{\bf Q}_{j}{\bf h}_{jli}\vert^{2} + \!\!\!\sum\limits_{i=1, i\ne k}^{K}\vert\frac{1}{N}\widehat{\mathbf{h}}_{jjk}^{H}{\bf Q}_{j}{\bf h}_{jji}\vert^{2}+\vert\frac{1}{N}\widehat{\mathbf{h}}_{jjk}^{H}{\bf Q}_{j}{\bf e}_{jjk}\vert^{2}+\frac{1}{{\rho^{\rm{ul}}}}\frac{1}{N}\widehat{\mathbf{h}}_{jjk}^{H}{\bf Q}_{j} \widehat{\mathbf{ h}}_{jjk}\left| \widehat{\mathbf{ H}}_{jj} \right.} \right\}}
\end{align}
\hrulefill
\end{figure*}
From the above results, it follows that
\begin{align}\nonumber
{\theta_{j}}{N}\!\!\!\!\sum \limits_{i=1,i\ne k}^{K} \!\!&\mathbb{E}\{\vert \frac{1}{N}\mathbf{h}_{jjk}^{H}\mathbf{\widehat h}_{jji}\vert^2\}\asymp \frac{1}{N} \overline\theta_{j}\!\!\!\!\!\sum \limits_{i=1,i\ne k}^{K}\!\!\!\!d_{jjk}\left( \!{\phi}_{jji} \!+\! \frac{1}{N}\overline {\mathbf{h}}_{jji}\overline {\mathbf{h}}_{jji}^{H}\right)\\  \label{A.21}&+ \frac{1}{N}{\phi}_{jji}\overline{\mathbf{h}}_{jjk}^{H}\overline{\mathbf{h}}_{jjk} + \frac{1}{N}\vert\overline {\mathbf{h}}_{jji}^{H}\overline{\mathbf{h}}_{jjk}\vert^{2}.
\end{align}
Putting \eqref{A.20} and \eqref{A.21} together yields:
\begin{align}\notag
 s_{jk}^{(I)} \asymp&\frac{1}{N}\sum \limits_{l=1}^{L} \sum\limits_{i=1, i\ne k}^{K}\overline \theta_{l}d_{ljk}\left( {\phi}_{lli} + \frac{1}{N}\overline {\mathbf{h}}_{lli}\overline {\mathbf{h}}_{lli}^{H}\right)  \\&+\frac{1}{N^{2}} \overline\theta_{j}\sum \limits_{i=1,i\ne k}^{K}{\phi}_{jji}\overline{\mathbf{h}}_{jjk}^{H}\overline{\mathbf{h}}_{jjk} + \vert\overline {\mathbf{h}}_{jji}^{H}\overline{\mathbf{h}}_{jjk}\vert^{2}
\end{align}
We now proceed considering the pilot contamination term $  s_{jk}^{(P)} $. Since $\widehat {\mathbf{h}}_{llk} $ depends on ${\mathbf{h}}_{ljk} $, we must proceed as follows. Recall that $\widehat {\mathbf{h}}_{llk} = \overline {\bf{h}}_{llk} +  \alpha_{llk}
  \left({\bf y}_{lk}^{\rm {tr}}  - \overline {\bf{h}}_{jjk} \right)$ and $\alpha_{llk} = {{d}_{llk}}{(\frac{1}{\rho^{\rm{tr}}} + \sum\nolimits_{n=1}^L   {d}_{lnk})^{-1}}$. Rewrite ${\widehat {\mathbf{h}}_{llk}}$ as follows 
\begin{align}\notag
 {\widehat {\mathbf{h}}}_{llk} &= \overline {\bf{h}}_{llk} +  \alpha_{llk}
  \left({\bf y}_{lk}^{\rm {tr}} -{\bf{h}}_{ljk}+{\bf{h}}_{ljk} - \overline {\bf{h}}_{jjk} \right) \\&= \widehat {\widehat {\mathbf{h}}}_{llk} + \alpha_{llk} {\bf{h}}_{ljk}\label{barbar_{h}llk}
\end{align}
with $\widehat {\widehat {\mathbf{h}}}_{llk} = \overline {\bf{h}}_{llk} +  \alpha_{llk}
  \left({\bf y}_{jk}^{\rm {tr}} -{\bf{h}}_{ljk}- \overline {\bf{h}}_{jjk} \right) =  {\widehat {\mathbf{h}}}_{llk} - \alpha_{llk} {\bf{h}}_{ljk}$. Using \eqref{barbar_{h}llk} we may write
\begin{align}\notag
\frac{1}{N^{2}}\vert \mathbf{h}_{ljk}^{H}\widehat {\mathbf{h}}_{llk}\vert^2&  =\alpha_{llk}^{2}\left\vert \frac{1}{N}\mathbf{h}_{ljk}^{H} {\bf{h}}_{ljk}\right\vert^2+\left\vert \frac{1}{N}\mathbf{h}_{ljk}^{H}\widehat {\widehat {\mathbf{h}}}_{llk}\right\vert^2 \\&+2\alpha_{llk}\Re e\left\{ \frac{1}{N^{2}}\widehat {\widehat {\mathbf{h}}}_{llk}\mathbf{h}_{ljk}\mathbf{h}_{ljk}^{H}\mathbf{h}_{ljk}\right\}.
\end{align}
Observe that $
\alpha_{llk}^{2}\left\vert \frac{1}{N}\mathbf{h}_{ljk}^{H} {\bf{h}}_{ljk}\right\vert^2 \asymp \alpha_{llk}^{2}d_{ljk}^{2} = \phi_{ljk}^{2}.$
Since $\widehat {\widehat {\mathbf{h}}}_{llk}$ is independent of $\mathbf{h}_{ljk}$, we have that$
| \frac{1}{N}\mathbf{h}_{ljk}^{H}\widehat {\widehat {\mathbf{h}}}_{llk}|^2 \asymp \frac{1}{N}d_{ljk}\big(\frac{1}{N}{\widehat {\widehat {\mathbf{h}}}}_{llk}^{H}\widehat {\widehat {\mathbf{h}}}_{llk}\big)$
with
\begin{align}\notag
\frac{1}{N}\widehat {\widehat {\mathbf{h}}}_{llk}^{H}\widehat {\widehat {\mathbf{h}}}_{llk}  &= \frac{1}{N}\widehat {{\mathbf{h}}}_{llk}^{H} {\widehat {\mathbf{h}}}_{llk} - \alpha_{llk} \frac{1}{N}{\bf{h}}_{ljk}^{H}{\widehat {\mathbf{h}}}_{llk} \\&- \alpha_{llk}\frac{1}{N} {\widehat {\mathbf{h}}}_{llk}^{H}{\bf{h}}_{ljk} + \alpha_{llk}^{2}\frac{1}{N}{\bf{h}}_{ljk}^{H}{\bf{h}}_{ljk}.
\end{align}
Observe now that $\frac{1}{N}\widehat {{\mathbf{h}}}_{llk}^{H} {\widehat {\mathbf{h}}}_{llk} \asymp {\phi}_{llk} +\frac{1}{N}\overline {\bf{h}}_{llk}^{H}\overline {\bf{h}}_{llk}$ whereas
\begin{align}
\frac{1}{N}{\bf{h}}_{ljk}^{H}{\widehat {\mathbf{h}}}_{llk} \asymp  \alpha_{llk}d_{ljk} \end{align}
and $\frac{1}{N}{\bf{h}}_{ljk}^{H}{\bf{h}}_{ljk} \asymp d_{ljk}.$
Putting all the above results together yields
        \begin{align}
 \!\!\!\! s_{jk}^{(P)} \asymp \!\!\sum \limits_{l=1,l\ne j}^{L} \!\!\overline \theta_{l} \phi_{ljk}^{2} \!+ \!\frac{1}{N}\!\!\!\sum \limits_{l=1,l\ne j}^{L}\!\! \overline \theta_{l}  d_{ljk} \left( {\phi}_{llk} \!+\! \frac{1}{N}  \overline {\bf{h}}_{llk}\overline {\bf{h}}_{llk}^{H}\right)
\end{align} 
where we have neglected the term $\alpha_{llk}d_{ljk}^{2}$, which appears only $L-1$ times.
Combining the above results together yields
\begin{align}\notag
   &\!\!\!\!\!\!s_{jk}^{(I)} +  s_{jk}^{(P)}\asymp\frac{1}{N}\sum \limits_{l=1}^{L} \sum\limits_{i=1}^{K}\overline \theta_{l}d_{ljk}\left( {\phi}_{lli} + \frac{1}{N}\overline {\mathbf{h}}_{lli}\overline {\mathbf{h}}_{lli}^{H}\right) \\ &\!\!\!\!\!\!+ \frac{1}{N^{2}} \overline\theta_{j}\!\!\!\!\sum \limits_{i=1,i\ne k}^{K}\!\!\left({\phi}_{jji}\overline{\mathbf{h}}_{jjk}^{H}\overline{\mathbf{h}}_{jjk} + \vert\overline {\mathbf{h}}_{jji}^{H}\overline{\mathbf{h}}_{jjk}\vert^{2}\right) + \sum\limits_{l\ne j}\overline\theta_{l} \phi_{ljk}^{2}
\end{align}
where we have added $\frac{1}{N}\overline \theta_{j}d_{jjk}\left( {\phi}_{jjk} + \frac{1}{N}\overline {\mathbf{h}}_{jjk}\overline {\mathbf{h}}_{jjk}^{H}\right)$, which is negligible for large $N$. We are only left with $\mathrm{var}\left[\frac{1}{N}\mathbf{h}_{jjk}^{H}{\widehat {\mathbf{h}}}_{jjk}\right]$, which can be rewritten as:
\begin{align}
\!\!\!\!\!\!\!\!\mathrm{var}\!\!\left[\frac{1}{N}\mathbf{h}_{jjk}^{H}{\widehat {\mathbf{h}}}_{jjk}\right] \!\!=\! \mathbb{E}\!\left\{\!\left|\frac{1}{N}\mathbf{h}_{jjk}^{H}{\widehat {\mathbf{h}}}_{jjk}\right|^{2}\!\right\} \!\!-\! \left|\mathbb{E}\!\left\{\frac{1}{N}\mathbf{h}_{jjk}^{H}{\widehat {\mathbf{h}}}_{jjk}\!\!\right\}\right|^{2}\!\!\!\!
\end{align}
from which it is easily follows that $\mathrm{var}\left[\frac{1}{N}\mathbf{h}_{jjk}^{H}{\widehat {\mathbf{h}}}_{jjk}\right]\asymp 0$ since $\frac{1}{N}\mathbf{h}_{jjk}^{H}{\widehat {\mathbf{h}}}_{jjk} \asymp{\phi}_{jjk} +\frac{1}{N}\overline {\bf{h}}_{jjk}^{H}\overline {\bf{h}}_{jjk}.$

	\section*{Appendix B \\ Asymptotic analysis of S-MMSE}
We start by dividing the numerator and denominator of $\gamma_{jk}^{{\rm{ul}}}$ by $1/N$. Then, we obtain \eqref{gamma_jk_App1} on the top of this page.
\smallskip
\subsubsection*{B.1) Preliminaries}
To begin with, we define ${\bf Q}_{jk}$ as:
\begin{align}
{\bf Q}_{jk}=\left(\frac{1}{N}\sum_{i\neq k}\widehat{\bf h}_{jji}\widehat{\bf h}_{jji}^{H}+\lambda_j{\bf I}_N\right)^{-1}.
%\tilde{\bf Q}_j&=\left(\widehat{\bf H}_{jj}^{H}\widehat{\bf H}_{jj}+\lambda_j{\bf I}_K\right)^{-1}
\end{align} 
Then, the following relations hold true:
\begin{align}
[\tilde{\bf Q}_j]_{kk}&=\frac{1}{\lambda\left(1+\widehat{\bf h}_{jjk}^{H}{\bf Q}_j\widehat{\bf h}_{jjk}\right)}\label{eq:qtilde}\\
{\bf Q}_j&={\bf Q}_{jk}-\lambda_j[\tilde{\bf Q}_j]_{kk}{\bf Q}_{jk}\frac{1}{N}\widehat{\bf h}_{jjk}\widehat{\bf h}_{jjk}^{H}{\bf Q}_{jk}\\
{\bf Q}_j\widehat{\bf h}_{jjk}&=\frac{{\bf Q}_{jk}\widehat{\bf h}_{jjk}}{1+\widehat{\bf h}_{jjk}^{H}{\bf Q}_j\widehat{\bf h}_{jjk}}=\lambda_j[\tilde{\bf Q}_j]_{kk}{\bf Q}_{jk}\widehat{\bf h}_{jjk}.
\end{align}
These relations will be extensively used in the asymptotic calculations of the SINR, where the replacement by ${\bf Q}_{jk}$ allows to ensure the dependence of   the resolvent matrix from $\widehat{\bf h}_{jjk}$. However, the direct replacement of the deterministic equivalents associated with ${\bf Q}_{jk}$ by those with ${\bf Q}_j$ cannot be performed in all cases - since we are dealing with non-centered random variables - especially when quadratic forms are involved. Some technical derivations are required to express all the terms in terms of only the deterministic equivalents of ${\bf Q}_j$. 

We shall also define the deterministic equivalents associated with ${\bf Q}_{jk}$.
Let $\delta_{jk}$ and $\tilde{\delta}_{jk}$ be the solutions to the following set of equations:
\begin{align*}
\delta_{jk}&=\frac{1}{N}\tr \left(\lambda_j(1+\tilde{\delta}_{jk}){\bf I}_N+\frac{1}{N}\overline{\bf H}_{jj}^{[k]}\left({\bf I}_{K-1}+\delta_{jk}\boldsymbol{\Phi}_{jj}^{[k]}\right)^{-1}\overline{\bf H}_{jj}^{[k]^{H}}\right)^{-1}\\
\tilde{\delta}_{jk}&=\frac{1}{N}\tr \boldsymbol{\Phi}_{jj}^{[k]}\left(\lambda_j({\bf I}_{K-1}+\delta_{jk}\boldsymbol{\Phi}_{jj}^{[k]})+\frac{1}{N}\frac{\overline{\bf H}_{jj}^{[k]^{H}}\overline{\bf H}_{jj}^{[k]}}{1+\tilde{\delta}_{jk}}\right)^{-1}
\end{align*}
where $\boldsymbol{\Phi}_{jj}^{[k]}$ is  $\boldsymbol{\Phi}_{jj}$  with the $k$-th row and $k$-th column removed and $\overline{\bf H}_{jj}^{[k]}$ is $\overline{\bf H}_{jj}$ after removal of the $k$-th column.
The following relations are also needed. For any $N\times N$ matrix, ${\bf B}$ and $N\times 1$ vector ${\bf b}$ with bounded norms, we have that
\begin{align}
&{\frac{1}{N}}\tr ({\bf B}{\bf T}_{jk})=\frac{1}{N}\tr ({\bf B}{\bf T}_j) +O(N^{-1})\label{eq:trace}\\
&\frac{1}{\sqrt{N}}{\bf b}^{H}{\bf T}_{jk}\overline{\bf h}_{jjk}=\frac{1}{\sqrt{N}}\frac{{\bf b}^{H}{\bf T}_j\overline{\bf h}_{jjk}}{\lambda_j\left[\widetilde{\bf T}_j\right]_{kk}(1+\phi_{jjk}\delta_j)} +O(N^{-1})\label{eq:q_form}\\
&[\tilde{\bf T}_j]_{kk}=\frac{1}{\lambda_j\left(1+\frac{1}{N}\overline{\bf h}_{jjk}^{H}{\bf T}_{jk}\overline{\bf h}_{jjk}+\tilde{\delta}_j\phi_{jjk}\right)}+O(N^{-1}).
\end{align}
The above relations will be needed to express deterministic equivalents involving ${\bf T}_{jk}$ in terms of ${\bf T}_j$ instead. We will also require that:
\begin{align}\notag
(1+\delta_j\phi_{jjk})^{-2}&\frac{1}{N}\overline{\bf h}_{jjk}^{H}{\bf T}_j\overline{\bf h}_{jjk}=\\&(1+\delta_j\phi_{jjk})^{-1}-\lambda_j[\tilde{\bf T}_j]_{kk}.
\label{eq:sim}
\end{align} 

\begin{figure*}
\begin{align}\notag
\lambda_j^2[\widetilde{\bf T}_j]_{kk}^2\sum_{i\neq k}\mathbb{E}\left\{\left|\frac{1}{N}\overline {\bf{h}}_{jjk}^{H}{\bf Q}_{jk}{\bf h}_{jli}\right|^2\right\}&=\frac{1}{(1+\delta_j\phi_{jjk})^2}\sum_{i\neq k}\mathbb{E}\left\{\left|\frac{1}{N}\overline {\bf{h}}_{jjk}^{H}{\bf Q}_j{\bf h}_{jli}\right|^2\right\}\\&-\phi_{jjk} \left(\frac{\frac{1}{N}\overline {\bf{h}}_{jjk}^{H}{\bf T}_j\overline {\bf{h}}_{jjk}}{(1+\delta_j\phi_{jjk})^2}\right)^2 \sum_{i\neq k} \frac{1}{N^2}\mathbb{E}\left\{{\bf h}_{jli}^{H}{\bf Q}_{jk}^2{\bf h}_{jli}\right\}+o(1).
\label{eq:f}
\end{align}
\hrulefill
\begin{align}\label{eq:outer}
s_{jk,{\rm outer}}^{(I)}&=\sum_{l\neq j}\sum_{i\neq k}\frac{\mathbb{E}\left\{\left|\frac{1}{N}\overline {\bf{h}}_{jjk}^{H}{\bf Q}_j{\bf h}_{jli}\right|^2\right\}}{(1+\delta_j\phi_{jjk})^2}+\phi_{jjk}\left(\lambda_j^{2}[\widetilde{\bf T}_j]_{kk}^2-\left(\frac{\frac{1}{N}\overline {\bf{h}}_{jjk}^{H}{\bf T}_j\overline {\bf{h}}_{jjk}}{(1+\delta_j\phi_{jjk})^2}\right)^2\right)\sum_{i\neq k} \frac{1}{N^2}\mathbb{E}\left\{{\bf h}_{jli}^{H}{\bf Q}_{j}^2{\bf h}_{jli}\right\} +o(1)
\end{align}
\hrulefill
\end{figure*}
\smallskip
\subsubsection*{B.2) Signal power}
Applying the matrix inversion lemma we may write:
\begin{align}\label{E1}
	\frac{1}{N}\widehat{\bf h}_{jjk}^H\mathbf{ Q}_{j}\mathbf{\widehat h}_{jjk}  &=\lambda_j[\tilde{\bf Q}_j]_{kk} \frac{1}{N}\widehat{\bf h}_{jjk}^{H}{\bf Q}_{jk}\widehat{\bf h}_{jjk}
\end{align}
 From \eqref{eq:qtilde}, $\frac{1}{N}\widehat{\bf h}_{jjk}{\bf Q}_{jk}\widehat{\bf h}_{jjk}=\frac{1}{\lambda_j [\tilde{\bf Q}_j]_{kk}}-1$. Using the fact that $[\tilde{\bf Q}_j]_{kk}\asymp[\widetilde{\bf T}_j]_{kk}$, we ultimately obtain:
\begin{align}
\frac{1}{N}\widehat{\bf h}_{jjk}^{H}{\bf Q}_j\widehat{\bf h}_{jjk}\asymp\left(1-\lambda_j[\widetilde{\bf T}_j]_{kk}\right)
\end{align}
and from the continuous mapping Theorem~\cite{Wagner12},   $|\frac{1}{N}\widehat{\bf h}_{jjk}^{H}{\bf Q}_j\widehat{\bf h}_{jjk}|^2\asymp(1-\lambda_j[\widetilde{\bf T}_j]_{kk})^2$.

\smallskip
 \subsubsection*{B.3) Interference power} In the asymptotic regime, the conditional expectation with respect to $\widehat{\bf H}_{jj}$   can be replaced by the expectation. We thus consider computing the following equivalent interference $ s_{jk}= s_{jk}^{(I)}  +  s_{jk}^{(P)} + o(1)$ where
      \begin{align}
      s_{jk}^{(I)}=\sum_{l=1}^{L} \sum_{i\neq k}\mathbb{E}\left\{\left|\frac{1}{N}\widehat{\bf h}_{jjk}^{H}{\bf Q}_j{\bf h}_{jli}\right|^2 \right\}
  \end{align}
  accounts for the intracell and intercell interference
 and 
        \begin{align}
s_{jk}^{(P)}=\sum_{l\neq j}\mathbb{E}\left\{\left|\frac{1}{N}\widehat{\bf h}_{jjk}^{H}{\bf Q}_j{\bf h}_{jlk}\right|^2\right\}
  \end{align}
  is due to pilot-sharing UEs. Let's start with $ s_{jk}^{(P)} $ for which it easily follows that:
\begin{align}
s_{jk}^{(P)}=\sum_{l\neq j}\left(\phi_{jlk}\delta_j\lambda_j[\widetilde{\bf T}_j]_{kk}\right)^2+o(1).
  \end{align}
The term $ s_{jk}^{(I)}$ is decomposed as
       \begin{align}\notag
s_{jk}^{(I)} &=\sum_{l\neq j} \sum_{i\neq k}\mathbb{E}\left\{\left|\frac{1}{N}\widehat{\bf h}_{jjk}^{H}{\bf Q}_j{\bf h}_{jli}\right|^2\right\} \\\notag&\quad\quad+ \sum_{i\neq k}\mathbb{E}\left\{\left|\frac{1}{N}\widehat{\bf h}_{jjk}^{H}{\bf Q}_j{\bf h}_{jji}\right|^2\right\} \\& \triangleq s_{jk,{\rm outer}}^{(I)} + s_{jk,{\rm inner}}^{(I)}
\end{align}
where $s_{jk,{\rm outer}}^{(I)}$ and $s_{jk,{\rm inner}}^{(I)}$ represent the inter-cell and intra-cell interference, respectively.
\begin{figure*}
\begin{align}
s_{jk,{\rm inner}}^{(I)}&=\sum_{i\neq k}\frac{\mathbb{E}\left\{\left|\frac{1}{N}\overline {\bf{h}}_{jjk}^{H}{\bf Q}_j{\bf h}_{jji}\right|^2\right\}}{(1+\delta_j\phi_{jjk})^2}+\phi_{jjk}\left(\lambda_j^{2}[\widetilde{\bf T}_j]_{kk}^2-\left(\frac{\frac{1}{N}\overline {\bf{h}}_{jjk}^{H}{\bf T}_j\overline {\bf{h}}_{jjk}}{(1+\delta_j\phi_{jjk})^2}\right)^2\right)\sum_{i\neq k}\frac{1}{N^2}\mathbb{E}\left\{{\bf h}_{jji}^{H}{\bf Q}_{j}^2{\bf h}_{jji}\right\} +o(1)
\label{eq:inner}
\end{align}
\hrulefill
\end{figure*}

We start with $s_{jk,{\rm outer}}^{(I)}$. By using ${\bf Q}_j\widehat{\bf h}_{jjk}=\lambda_j[\tilde{\bf Q}_j]_{kk} {\bf Q}_{jk}\widehat{\bf h}_{jjk}$ and by taking the expectation over $\widehat{\bf h}_{jjk}$ yields
\begin{align}\notag
&\!\!\!\!\!\!\!s_{jk,{\rm outer}}^{(I)}=\lambda_j^2[\widetilde{\bf T}_j]_{kk}^2\sum_{l\neq j}\sum_{i\neq k}\mathbb{E} \left\{\left|\frac{1}{N}\overline {\bf{h}}_{jjk}^{H}{\bf Q}_{jk}{\bf h}_{jli}\right|^2\right\} \\&+\lambda_j^2\phi_{jjk}[\widetilde{\bf T}_j]_{kk}^2\sum_{l\neq j} \sum_{i\neq k} \mathbb{E}\left\{\frac{1}{N^2}{\bf h}_{jli}^{H}{\bf Q}_{jk}^2{\bf h}_{jli}\right\}+o(1).\!\!\!
\label{eq:outer.10}
\end{align}
In order to get simplified expressions, we need to work with ${\bf Q}_j$ instead of ${\bf Q}_{jk}$. As mentioned before, a direct replacement of ${\bf Q}_j$ with ${\bf Q}_{jk}$ is not allowed since it induces a non-vanishing error that needs to be evaluated beforehand. Hence, we use the following identity
       \begin{align}
{\bf Q}_j={\bf Q}_{jk}-\lambda_j[\tilde{\bf Q}_j]_{kk}{\bf Q}_{jk} \frac{1}{N}\widehat{\bf h}_{jjk}\widehat{\bf h}_{jjk}^{H}{\bf Q}_{jk}
       \end{align}
to obtain
\begin{align}\notag
&\sum_{i\neq k} \mathbb{E}\left\{\left|\frac{1}{N}\overline {\bf{h}}_{jjk}^{H}{\bf Q}_j{\bf h}_{jli}\right|^2\right\}=\sum_{i\neq k} \mathbb{E}\left\{\left|\frac{1}{N}\overline {\bf{h}}_{jjk}^{H}{\bf Q}_{jk}{\bf h}_{jli}\right|^2\right\} \\\notag&+\sum_{i\neq k} \lambda_j^2[\widetilde{\bf T}_j]_{kk}^2\frac{1}{N^4}\mathbb{E}\left\{\left|\overline {\bf{h}}_{jjk}^{H}{\bf Q}_{jk}\widehat{\bf h}_{jjk}\right|^2\left|{\bf h}_{jli}^{H}{\bf Q}_{jk}\widehat{\bf h}_{jjk}\right|^2\right\} \\ \notag&-\sum_{i\neq k}\lambda_j[\widetilde{\bf T}_j]_{kk}\frac{1}{N^3}\mathbb{E}\left\{\overline {\bf{h}}_{jjk}^{H}{\bf Q}_{jk}\widehat{\bf h}_{jjk}\widehat{\bf h}_{jjk}^{H}{\bf Q}_{jk}{\bf h}_{jli}\widehat{\bf h}_{jli}^{H}{\bf Q}_{jk}\overline {\bf{h}}_{jjk}\right\}\\\notag
&-\sum_{i\neq k}\lambda_j[\widetilde{\bf T}_j]_{kk}\frac{1}{N^3}\mathbb{E}\left\{\overline {\bf{h}}_{jjk}^{H}{\bf Q}_{jk}{\bf h}_{jli}{\bf h}_{jli}^{H}{\bf Q}_{jk}\widehat{\bf h}_{jjk}\widehat{\bf h}_{jjk}^{H}{\bf Q}_{jk}\overline {\bf{h}}_{jjk}\right\}\\&+o(1)
\end{align}
which reduces to
\begin{align}\notag
&\sum_{i\neq k} \mathbb{E}\left\{\left|\frac{1}{N}\overline {\bf{h}}_{jjk}^{H}{\bf Q}_j{\bf h}_{jli}\right|^2\right\}
=\notag\sum_{i\neq k} \mathbb{E}\left\{\left|\frac{1}{N}\overline{\bf h}_{jjk}^{H}{\bf Q}_{jk}{\bf h}_{jli}\right|^2\right\} \\\notag&-2\lambda_j[\tilde{\bf T}_j]_{kk}\frac{1}{N}\overline{\bf h}_{jjk}^{H}{\bf T}_{jk}\overline{\bf h}_{jjk} \mathbb{E}\left\{\left|\frac{1}{N}\overline{\bf h}_{jjk}^{H}{\bf Q}_{jk}{\bf h}_{jli}\right|^2\right\}\\
&+\notag\lambda_j^2[\tilde{\bf T}_j]_{kk}^2\!\!\left(\frac{1}{N}\overline{\bf h}_{jjk}^{H}{\bf T}_{jk}\overline{\bf h}_{jjk}\right)^2\!\!\Bigg(\!\!\sum_{i\neq k}\mathbb{E}\left\{\left|\frac{1}{N}\overline{\bf h}_{jjk}^{H}{\bf Q}_{jk}{\bf h}_{jli}\right|^2\right\}\!+ \\\notag&+\phi_{jjk}\sum_{i\neq k}\mathbb{E}\left\{\left|\frac{1}{N}{\bf h}_{jli}^{H}{\bf Q}_{jk}^2{\bf h}_{jli}\right|^2\right\}\Bigg)+o(1)\\
&=\notag\left(1-\lambda_j[\widetilde{\bf T}_j]_{kk}\frac{1}{N}\overline{\bf h}_{jjk}^{H}{\bf T}_{jk}\overline{\bf h}_{jjk}\right)^2\!\!\!\!\sum_{i\neq k}\mathbb{E}\left\{\left|\frac{1}{N}\overline{\bf h}_{jjk}^{H}{\bf Q}_{jk}{\bf h}_{jli}\right|^2\right\}\\\notag
&+\lambda_j^2[\widetilde{\bf T}_j]_{kk}^2\left(\frac{1}{N}\overline{\bf h}_{jjk}^{H}{\bf T}_{jk}\overline{\bf h}_{jjk}\right)^2\sum_{i\neq k} \mathbb{E}\left\{\left|\frac{1}{N}{\bf h}_{jli}^{H}{\bf Q}_{jk}^2{\bf h}_{jli}\right|^2\right\}\\&+o(1).
\end{align}
By using \eqref{eq:q_form} and \eqref{eq:sim}, we obtain
\eqref{eq:f} on the top of this page.
By plugging \eqref{eq:f} into \eqref{eq:outer} and replacing ${\bf Q}_{jk}$ with ${\bf Q}_j$ in the second term (up to a vanishing error), we obtain \eqref{eq:outer} on the top of this page. Similarly, we can show that $s_{jk,{\rm inner}}^{(I)}$ is given by \eqref{eq:inner} on the top of next page.

Consider the term $\sum_{i\neq k}\mathbb{E}\{|\frac{1}{N}\overline {\bf{h}}_{jjk}^{H}{\bf Q}_j{\bf h}_{jli}|\}^2$ in \eqref{eq:outer} for $l\neq j$. By using the identity
\begin{equation}
{\bf Q}_{j}={\bf Q}_{ji}-\lambda_j[\tilde{\bf Q}_j]_{ii}\frac{1}{N}{\bf Q}_{ji}\widehat{\bf h}_{jji}\widehat{\bf h}_{jji}^{H}{\bf Q}_{ji}
\label{eq:inversion_i}
\end{equation}
and replacing $[\tilde{\bf Q}_j]_{ii}$ with $[\widetilde{\bf T}_j]_{ii}$ and $\frac{1}{N}{\widehat{\mathbf{h}}}_{jji}^{H}{\bf Q}_{ji}{\bf h}_{jli}$ with $\delta_j\phi_{jli}$, we obtain:
\begin{align}
&\notag\sum_{i\neq k} \mathbb{E}\left|\overline{\bf h}_{jjk}^{H}{\bf Q}_j{\bf h}_{jli}\right|^2=\sum_{i\neq k} \mathbb{E}\left\{\left|\frac{1}{N}\overline{\bf h}_{jjk}^{H}{\bf Q}_{ji}{\bf h}_{jli}\right|^2\right\} \\\notag&-\sum_{i\neq k} \lambda_j[\widetilde{\bf T}_j]_{ii}\phi_{jli}\delta_j \frac{1}{N^2}\mathbb{E}\left\{\overline{\bf h}_{jjk}^{H}{\bf Q}_{ji}{\widehat{\mathbf{h}}}_{jji}{\bf h}_{jli}^{H}{\bf Q}_{ji}\overline{\bf h}_{jjk}\right\}\\
&\notag-\sum_{i\neq k} \lambda_j[\tilde{\bf T}_j]_{ii}\phi_{jli}\delta_j \frac{1}{N^2}\mathbb{E}\left\{\overline{\bf h}_{jjk}^{H}{\bf Q}_{ji}{\bf h}_{jli}{\widehat{\mathbf{h}}}_{jji}^{H}{\bf Q}_{ji}\overline{\bf h}_{jjk}\right\}\\\notag&+\sum_{i\neq k} \lambda_j^2[\tilde{\bf T}_j]_{ii}^2\phi_{jli}^2\delta_j^2 \frac{1}{N^2}\mathbb{E}\left\{\overline{\bf h}_{jjk}^{H}{\bf Q}_{ji}{\widehat{\mathbf{h}}}_{jji}{\widehat{\mathbf{h}}}_{jji}^{H}{\bf Q}_{ji}\overline{\bf h}_{jjk}\right\}\\
&+o(1).
\end{align}
Computing the expectation over ${\bf h}_{jli}$ and ${\widehat{\mathbf{h}}}_{jji}$ yields
\begin{align}
&\notag\sum_{i\neq k} \mathbb{E}\left|\overline{\bf h}_{jjk}^{H}{\bf Q}_j{\bf h}_{jli}\right|^2=\\\notag&=\sum_{i\neq k}\left(d_{jli}-2\lambda_j[\widetilde{\bf T}_j]_{ii}\phi_{jli}^2\delta_j+\lambda_j^2[\widetilde{\bf T}_j]_{ii}^2\phi_{jli}^2\delta_j^2\phi_{jji}\right) \times\\\notag&\quad\quad\frac{1}{N^2} \mathbb{E}\left\{\overline{\bf h}_{jjk}^{H}{\bf Q}_{ji}^2\overline{\bf h}_{jjk}\right\}+\\\notag
&+\sum_{i\neq k}\lambda_j^2[\widetilde{\bf T}_j]_{ii}^2\phi_{jli}^2\delta_j^2\frac{1}{N^2}\mathbb{E}\left\{\overline{\bf h}_{jjk}^{H}{\bf Q}_{ji}\overline{\bf h}_{jji}\overline{\bf h}_{jji}^{H}{\bf Q}_{ji}\overline{\bf h}_{jjk}\right\}\\&+o(1).
\end{align}
In the first term of the above equation, ${\bf Q}_{ji}$ can be replaced by ${\bf Q}_j$ up to a vanishing error. However, this is not the case of the second term, for which the replacement of ${\bf Q}_{ji}$ with ${\bf Q}_j$ is not allowed. To handle this term, we propose to work out the 
$
\frac{1}{N^2}\sum_{i\neq k} \phi_{jli}^2\delta_j^2(1+\delta_j\phi_{jji})^{-2}\mathbb{E}\{\overline{\bf h}_{jjk}^{H}{\bf Q}_j\overline{\bf h}_{jji}\overline{\bf h}_{jji}^{H}{\bf Q}_j\overline{\bf h}_{jjk}
\}$. Using the relation ${\bf Q}_{j}={\bf Q}_{ji}-\lambda_j[\widetilde{\bf Q}_j]_{ii} {\bf Q}_{ji}{\widehat{\mathbf{h}}}_{jji}{\widehat{\mathbf{h}}}_{jji}^{H}{\bf Q}_{ji}$, we obtain after some simplification:
\begin{align}\notag
&\frac{1}{N^2}\sum_{i\neq k} \phi_{jli}^2\delta_j^2(1+\delta_j\phi_{jji})^{-2}\mathbb{E}\{\overline{\bf h}_{jjk}^{H}{\bf Q}_j\overline{\bf h}_{jji}\overline{\bf h}_{jji}^{H}{\bf Q}_j\overline{\bf h}_{jjk}\}=\\\notag&\frac{1}{N^2}\sum_{i\neq k} \lambda_j^2[\widetilde{\bf T}_j]_{ii}^2 \phi_{jli}^2\delta_j^2 \mathbb{E}\left\{\overline{\bf h}_{jjk}^{H}{\bf Q}_{ji}\overline{\bf h}_{jji}\overline{\bf h}_{jji}^{H}{\bf Q}_{ji}\overline{\bf h}_{jjk}\right\}\\
&+\frac{1}{N^2}\sum_{i\neq k} \phi_{jji} \phi_{jli}^2\delta_j^2\frac{(\frac{1}{N}\overline{\bf h}_{jji}^{H}{\bf T}_j\overline{\bf h}_{jji})^2}{(1+\delta_j\phi_{jji})^4}+o(1).
\end{align} 
By using this identity, the first term in $s_{jk,{\rm outer}}^{(I)}$ simplifies to:
\begin{align}\notag
&\frac{1}{N^2}\sum_{i\neq k}\frac{\mathbb{E}\left\{\left|\frac{1}{N}\overline{\bf h}_{jjk}^{H}{\bf Q}_j{\bf h}_{jli}\right|^2\right\}}{(1+\delta_j\phi_{jjk})^2}\\\notag&=\frac{1}{N^2}\sum_{i\neq k} \phi_{jli}^2\delta_j^2\frac{\mathbb{E}\left\{\overline{\bf h}_{jjk}^{H}{\bf Q}_j\overline{\bf h}_{jji}\overline{\bf h}_{jji}^{H}{\bf Q}_j\overline{\bf h}_{jjk}\right\}}{(1+\delta_j\phi_{jjk})^2(1+\delta_j\phi_{jji})^2}\notag\\
&+\sum_{i \neq k} \frac{\mathbb{E}\left\{\frac{1}{N^2}\overline{\bf h}_{jjk}^{H}{\bf Q}_j^2\overline{\bf h}_{jjk}\right\}}{(1+\delta_j\phi_{jjk})^2} \left(\mu_{jli}-\tilde{\beta}_{jli}\right)+o(1)\label{eq:plug}
\end{align}
where $\mu_{jli}$ and $\gamma_{jli}$ are defined in \eqref{mu} and \eqref{gamma}, respectively, and 
\begin{align}\notag
&\tilde{\beta}_{jli}=\phi_{jli}^2\delta_j^2\lambda_j^2[\widetilde{\bf T}_j\boldsymbol{\Phi}_j\widetilde{\bf T}_j]_{ii}-\\&-\lambda_j^2\phi_{jli}^2\delta_j^2\phi_{jji}[\widetilde{\bf T}_j]_{ii}^2 +\phi_{jli}^2\delta_j^2\phi_{jji}\frac{\left|\frac{1}{N}\overline{\bf h}_{jji}^{H}{\bf T}_j\overline{\bf h}_{jji}\right|^2}{(1+\delta_j\phi_{jji})^4}
%\tilde{\gamma}_{jk}&=\frac{[\tilde{\bf T}_j\frac{1}{N}\overline{\bf H}_j^{H}\overline{\bf H}_j\tilde{\bf T}_j]_{kk}}{(1+\tilde{\delta}_j)^2}.
\end{align}
if $l\neq j$ 
and 
\begin{align}\notag
&\tilde{\beta}_{jjk}=\lambda_j^2[\widetilde{\bf T}_j\boldsymbol{\Phi}_{jj}\widetilde{\bf T}_j]_{kk}-\\&-\lambda_j^2\phi_{jjk}[\widetilde{\bf T}_j]_{kk}^2 +\phi_{jjk}\frac{\left|\frac{1}{N}\overline{\bf h}_{jjk}^{H}{\bf T}_j\overline{\bf h}_{jjk}\right|^2}{(1+\delta_j\phi_{jjk})^2}.
\end{align}
\begin{figure*}
\begin{align}\notag
\phi_{jjk}\left(\lambda_j^2[\widetilde{\bf T}_j]_{kk}^2-\left(\frac{\frac{1}{N}\overline {\bf{h}}_{jjk}^{H}{\bf T}_j\overline {\bf{h}}_{jjk}}{(1+\delta_j\phi_{jjk})^2}\right)^2\right)&\sum_{i\neq k}\frac{1}{N^2}\mathbb{E}\left\{{\bf h}_{jli}^{H}{\bf Q}_{j}^2{\bf h}_{jli}\right\}=\left(\lambda_j^2[\widetilde{\bf T}_j\boldsymbol{\Phi}_{jj}\widetilde{\bf T}_j]_{kk}-\tilde{\beta}_{jjk}\right)\frac{1}{N}\sum_{i=1}^{K}\frac{1-F_j}{\Delta_j}\gamma_{jli}\\
&+\left(\lambda_j^2[\widetilde{\bf T}_j\boldsymbol{\Phi}_{jj}\widetilde{\bf T}_j]_{kk}-\tilde{\beta}_{jjk}\right)\left(\frac{1}{N}\sum_{i=1}^{K}\mu_{jli}\overline{\nu}_j\right)+o(1).
\label{eq:sum_2}
\end{align}
\hrulefill
\begin{align}\notag
\sum_{i\neq k}\mathbb{E}\left\{\left|\frac{1}{N}\overline {\bf{h}}_{jjk}^{H}{\bf Q}_j{\bf h}_{jji}\right|^2\right\}&=\lambda_j^2\sum_{i\neq k}[\widetilde{\bf T}_j]_{ii}^2\mathbb{E}\left\{\left|\frac{1}{N}\overline {\bf{h}}_{jjk}^{H}{\bf Q}_{ji}\overline {\bf{h}}_{jji}\right|^2\right\} \\&+\frac{1}{N}\sum_{i\neq k} \left(d_{jji}-\phi_{jji}(1-\lambda_j^2[\widetilde{\bf T}_j]_{ii}^2)\right) \mathbb{E}\left\{\frac{1}{N}\overline {\bf{h}}_{jjk}^{H}{\bf Q}_j^2\overline {\bf{h}}_{jjk}\right\}+o(1).\label{eq:118}
\end{align}
\hrulefill
\begin{align}\notag
\sum_{i\neq k} \mathbb{E}\left|\frac{1}{N}\overline {\bf{h}}_{jjk}^{H}{\bf Q}_{j}{\bf h}_{jji}\right|^2&= \sum_{i\neq k} \frac{\mathbb{E}\left\{\left|\frac{1}{N}\overline {\bf{h}}_{jjk}^{H}{\bf Q}_j\overline {\bf{h}}_{jji}\right|^2\right\}}{(1+\delta_j\phi_{jji})^2}\\
&+\frac{1}{N}\sum_{i\neq k} \left(d_{jji}-\phi_{jji}\bigg(1+\lambda_j^2[\widetilde{\bf T}_j]_{ii}^2-\frac{(\frac{1}{N}\overline {\bf{h}}_{jji}^{H}{\bf T}_j\overline {\bf{h}}_{jji})^2}{(1+\delta_j\phi_{jji})^4}\bigg)\right) \mathbb{E}\left\{\frac{1}{N}\overline {\bf{h}}_{jjk}^{H}{\bf Q}_j^2\overline {\bf{h}}_{jjk}\right\}+o(1)\label{eq:B107new}
\end{align}
\hrulefill
\end{figure*}
From \cite[Corollaries 1 and 2]{Kammoun-18}, it follows that:
\begin{align}\notag\
&\frac{1}{N^2}\sum_{i\neq k} \phi_{jli}^2\delta_j^2 \frac{\mathbb{E}\left\{\overline{\bf h}_{jjk}^{H}{\bf Q}_j\overline{\bf h}_{jji}\overline{\bf h}_{jji}^{H}{\bf Q}_j\overline{\bf h}_{jjk}\right\}}{(1+\delta_j\phi_{jjk})^2(1+\delta_j\phi_{jji})^2}=\\\notag&\lambda_j^2[\widetilde{\bf T}_j{\rm diag}\left\{\phi_{jli}^2\delta_j\right\}_{i=1}^{K}\widetilde{\bf T}_j]_{kk}-\lambda_j^2[\widetilde{\bf T}_j]_{kk}^2\delta_j^2\phi_{jlk}^2+\\\notag
&+\tilde{\beta}_{jjk}\left(\frac{1-F_j}{\Delta_j}\frac{1}{N}\sum_{i\neq k}{\gamma}_{jli}+\overline{\nu}_j\frac{1}{N}\sum_{i=1}^{K}\tilde{\beta}_{jli}\right)\\&+{\gamma}_{jjk}\left(\frac{1-F_j}{\Delta_j}\frac{1}{N}\sum_{i=1}^{K}\tilde{\beta}_{jli}+\frac{\lambda_j^2\tilde{\vartheta}_j}{\Delta_j}\frac{1}{N}\sum_{i\neq k}{\gamma}_{jli}\right)\notag\\&+o(1)\label{plug2}
%&+\left(\lambda_j^2[\widetilde{\bf T}_j{\rm diag}\left\{\phi_{jli}^2\delta_j^2\right\}\widetilde{\bf T}_j]_{kk}-\lambda_j^2[\widetilde{\bf T}_j]_{kk}^2\delta_j^2\phi_{jlk}^2\right)
\end{align}
and
\begin{align}
\frac{\mathbb{E}\left\{\frac{1}{N}\overline{\bf h}_{jjk}^{H}{\bf Q}_j^2\overline{\bf h}_{jjk}\right\}}{(1+\delta_j\phi_{jjk})^2}=\frac{1-F_j}{\Delta_j} \gamma_{jjk} +\overline{\nu}_j \tilde{\beta}_{jjk}+o(1)\label{plug1}
\end{align}
where $\widetilde{\vartheta}_j=\frac{1}{N}\tr\left( \boldsymbol{\Phi}_{jj}\widetilde{\bf T}_j\right)^2$. By plugging \eqref{plug2} and \eqref{plug1} into \eqref{eq:plug}, we thus obtain:
\begin{align}\notag
&\sum_{i\neq k}\mathbb{E}\left\{\left|\frac{1}{N}\overline {\bf{h}}_{jjk}^{H}{\bf Q}_j{\bf h}_{jli}\right|^2\right\}=\lambda_j^2[\widetilde{\bf T}_j{\rm diag}\left\{\phi_{jli}^2\delta_j^2\right\}\widetilde{\bf T}_j]_{kk}\\\notag&-\lambda_j^2[\widetilde{\bf T}_j]_{kk}^2\delta_j^2\phi_{jlk}^2 +\tilde{\beta}_{jjk}\frac{1-F_j}{\Delta_j}\frac{1}{N}\sum_{i=1}^{K}\gamma_{jli}\\\notag&+{\gamma}_{jjk}\lambda_j^2\frac{\widetilde{\vartheta}_j}{\Delta_j}\frac{1}{N}\sum_{i=1}^{K}\gamma_{jli}\\
&+\left(\frac{1-F_j}{\Delta_j}\gamma_{jjk}+\overline{\nu}_j\tilde{\beta}_{jjk}\right) \left(\frac{1}{N}\sum_{i=1}^{K}\mu_{jli}\right)+o(1).\label{eq:sum_1}
\end{align}

\begin{figure*}
\begin{align}\notag
\phi_{jjk}\left(\lambda_j^2[\widetilde{\bf T}_j]_{kk}^2-\left(\frac{\frac{1}{N}\overline {\bf{h}}_{jjk}^{H}{\bf T}_j\overline {\bf{h}}_{jjk}}{(1+\delta_j\phi_{jjk})^2}\right)^2\right)&\sum_{i\neq k}\frac{1}{N^2}\mathbb{E}\left\{{\bf h}_{jji}^{H}{\bf Q}_j^2{\bf h}_{jji}\right\}=\\
& =\left(\lambda_j^2[\widetilde{\bf T}_j\boldsymbol{\Phi}_j\widetilde{\bf T}_j]_{kk}-\tilde{\beta}_{jjk}\right) \left(\frac{1-F_j}{\Delta_j}\frac{1}{N}\sum_{i=1}^{K}\overline{\nu}_j{\gamma}_{jji} +\overline{\nu}_j\mu_{jji}\right)+o(1).\label{eq:B108new}
\end{align}
\hrulefill
\end{figure*}

Consider now  the second term $\frac{1}{N^2}\mathbb{E}\left\{{\bf h}_{jli}^{H}{\bf Q}_{j}^2{\bf h}_{jli}\right\}$ in \eqref{eq:outer}. By using the results in \cite{Kammoun-18}, it can be proved that:
\begin{align}\notag
\sum_{i\neq k}\frac{1}{N^2}\mathbb{E}&\left\{{\bf h}_{jli}^{H}{\bf Q}_{j}^2{\bf h}_{jli}\right\}=\frac{1}{N}\sum_{i=1}^{K}\mu_{jli}\overline{\nu}_j\\&+\lambda_j^2[\widetilde{\bf T}_j\boldsymbol{\Phi}_j\widetilde{\bf T}_j]_{ii}\phi_{jli}^2\delta_j^2\overline{\nu}_j\frac{1-F_j}{\Delta_j}\gamma_{jli}+o(1)
\end{align}
from which \eqref{eq:sum_2} follows. By summing \eqref{eq:sum_1} and \eqref{eq:sum_2} for all $l\neq j$, we obtain:
\begin{align}\notag
&s_{jk,{\rm outer}}^{(I)}=\sum_{l\neq j}\lambda_j^2[\widetilde{\bf T}_j{\rm diag}\left\{\phi_{jli}^2\delta_j^2\right\}\widetilde{\bf T}_j]_{kk}-\lambda_j^2[\widetilde{\bf T}_j]_{kk}^2\delta_j^2\phi_{jlk}^2\\&+\frac{1}{N}\sum_{i=1}^K\gamma_{jli}\zeta_{jk}+\xi_{jk}\left(\frac{1}{N}\sum_{i=1}^{K}\mu_{jli}\right)+o(1).
\end{align}
Let's consider now the term $s_{jk,{\rm inner}}^{(I)}$. We begin with $\sum_{i\neq k}\mathbb{E}\{|\frac{1}{N}\overline {\bf{h}}_{jjk}^{H}{\bf Q}_j{\bf h}_{jji}\|^2\} $, which similar calculations allows to write explicitly as \eqref{eq:118}. The first term reduces to
\begin{align}\notag
&\lambda_j^2\sum_{i\neq k}[\widetilde{\bf T}_j]_{ii}^2\mathbb{E}\left\{\left|\frac{1}{N}\overline {\bf{h}}_{jjk}^{H}{\bf Q}_{ji}\overline {\bf{h}}_{jji}\right|^2\right\}=\\\notag&=\sum_{i\neq k} \frac{\mathbb{E}\left\{\left|\frac{1}{N}\overline {\bf{h}}_{jjk}^{H}{\bf Q}_j\overline {\bf{h}}_{jji}\right|^2\right\}}{(1+\delta_j\phi_{jji})^2}\\\notag&-\frac{1}{N}\sum_{i\neq k}\phi_{jji}\frac{\left(\frac{1}{N}\overline {\bf{h}}_{jji}^{H}{\bf T}_j\overline {\bf{h}}_{jji}\right)^2}{(1+\delta_j\phi_{jji})^4} \mathbb{E}\left\{\frac{1}{N}\overline {\bf{h}}_{jjk}^{H}{\bf Q}_j^2\overline {\bf{h}}_{jjk}\right\}\\&+o(1)
\end{align}
such that \eqref{eq:118} reduces to \eqref{eq:B107new}. Using standard calculations, we can show that:
\begin{align}\notag
&\sum_{i\neq k} \frac{\mathbb{E}\left\{\left|\frac{1}{N}\overline {\bf{h}}_{jjk}^{H}{\bf Q}_j{\bf h}_{jji}\right|^2\right\}}{(1+\delta_j\phi_{jji})^2}=\left(\lambda_j^2[\widetilde{\bf T}_j^2]_{kk}-\lambda_j^2[\widetilde{\bf T}_j]_{kk}^2\right)\\\notag&+\tilde{\beta}_{jjk}\frac{1-F_j}{\Delta_j}\frac{1}{N}\sum_{i=1}^{K}\gamma_{jji}+\gamma_{jjk}\frac{\lambda_j^2\widetilde{\vartheta}_j}{\Delta_j}\frac{1}{N}\sum_{i=1}^{K}\gamma_{jji}\\
&+\left(\frac{1-F_j}{\Delta_j}\gamma_{jjk}+\overline{\nu}_j\beta_{jk}\right)\left(\frac{1}{N}\sum_{i=1}^{K}\mu_{jji}\right)+o(1)
\end{align}
and
\begin{align}\notag
\mathbb{E}&\left\{\frac{1}{N}{\bf h}_{jji}^{H}{\bf Q}_j^2{\bf h}_{jji}\right\}=\frac{\vartheta_j}{\Delta_j}\left(d_{jji}-\phi_{jji}+\lambda_j^2[\widetilde{\bf T}_j\boldsymbol{\Phi}_j\widetilde{\bf T}_j]_{ii}\right) \\&+\frac{1-F_j}{\Delta_j} \frac{[\widetilde{\bf T}_j\frac{1}{N}\overline {\bf{H}}_{jj}^{H}\overline {\bf{H}}_{jj}\widetilde{\bf T}_j]_{ii}}{(1+\tilde{\delta}_j)^2}+o(1)
\end{align}
where $\vartheta_j=\frac{1}{N}\tr \left({\bf T}_j^2\right)$.
The second term in \eqref{eq:inner} can be approximated as in \eqref{eq:B108new} on the top of this page. Gathering all the above results together, we eventually obtain:
\begin{align}\notag
s_{jk,{\rm inner}}^{(I)}&=\lambda_j^2\big([\widetilde{\bf T}_j^2]_{kk}-[\widetilde{\bf T}_j]_{kk}^2\big)+\frac{1}{N}\sum_{i=1}^{K}\mu_{jji}\xi_{jk}\\&+\frac{1}{N}\sum_{i=1}^{K}\gamma_{jji}\zeta_{jk}+o(1). 
\end{align}

\begin{figure*}
\begin{align}\notag
\frac{\overline{s}}{\overline{\psi}}&=(\frac{d}{\phi}-1+\lambda^2\frac{N}{K\phi^2}\widetilde{\vartheta})\left(-1+\frac{1}{\Delta}\right) -\frac{F(d-\phi)}{\phi\Delta}-\frac{F(L-1)}{\Delta }\left(\frac{\alpha }{\phi}-2\lambda\alpha^2(1+\kappa)^2\delta^\star\frac{N}{K}\tilde{\delta}^\star\right)\\\label{eq:C115}
&+(L-1)\left(\frac{\alpha }{\phi}-2\lambda\alpha^2(1+\kappa)^2\delta^\star\frac{N}{K}\tilde{\delta}^\star+\alpha^2(1+\kappa)^2(\delta^\star)^2\lambda^2\frac{N}{K}\widetilde{\vartheta}\right)\left(-1+\frac{1}{\Delta}\right)
\end{align}
\hrulefill
\begin{align}\notag
\frac{\overline{s}}{\overline{\psi}}&=\left(\frac{1+\kappa-\nu}{\nu}+\lambda^2\left(\frac{N}{K}\right)^2\frac{(\tilde{\delta}^\star)^2}{\nu^2}(1+\kappa)^4\right)\left(-1+\frac{1}{\Delta}\right)-\frac{1+\kappa-\nu}{\nu^2}\frac{\kappa(1+\kappa)}{\Delta}\frac{N}{K}\frac{(\tilde{\delta}^\star)^2}{(1+\tilde{\delta}^\star)^2}\\\notag
&+(L-1)\left(\frac{\alpha(1+\kappa)^2}{\nu}-2\lambda\alpha^2(1+\kappa)^2\delta^\star\tilde{\delta}^\star\frac{N}{K}+\alpha^2(1+\kappa)^2(\delta^\star)^2\lambda^2\left(\frac{N}{K}\right)^2(\tilde{\delta}^\star)^2\right)\bigg(-1+\frac{1}{\Delta}\bigg)\\
&-\frac{(L-1)}{\Delta}\frac{\kappa(1+\kappa)}{\nu}\frac{N}{K}\frac{(\tilde{\delta}^\star)^2}{(1+\tilde{\delta}^\star)^2}\left(\frac{\alpha(1+\kappa)^2}{\nu}-2\lambda\alpha^2(1+\kappa)^2\delta^\star\tilde{\delta}^\star\frac{N}{K}\right)\label{eq:C:119new}
\end{align}
\hrulefill
\end{figure*}

    \section*{  Appendix C \\ Proof of Corollary \ref{corollary6}}\label{AppendixD}
If the channel is modeled as in \eqref{simplified_{channel}} and $\frac{1}{N}\overline {\bf{H}}_{jj}^{H}\overline {\bf{H}}_{jj}=\frac{\kappa}{1+\kappa}{\bf I}_{K}$, then from \eqref{lambda_j} we obtain that 
\begin{align}
\lambda_j=\lambda = \frac{K}{N}\Big(\alpha (L-1) + \frac{1}{1+\kappa} - \phi\Big)
\end{align}
and ${{\boldsymbol{ \Phi}}}_{jj} = \phi {\bf I}_K$ with $\phi_{jjk}\mathop  = \limits^{(a)}\phi= \frac{\nu}{(1+\kappa)^2}$, where $(a)$ follows from \eqref{phi_{jlk}_case_study}. Also, it turns out that ${\delta}_j = {\delta}$ and $\widetilde{\delta}_j = \widetilde{\delta}$ with
\begin{align}\label{delta_j_simplified}
\widetilde{\delta} =\phi \frac{K}{N}\left({\lambda+ \lambda\widetilde{\delta} + \phi\Big(1-\frac{K}{N}\Big) + \frac{\kappa}{1+\kappa}\frac{1}{1 +\widetilde{\delta}}}\right)^{-1}
\end{align}
where we have used that (after some simple calculus) $\phi{\delta} = \widetilde{\delta} + \phi \frac{1-K/N}{\lambda}$.
%\begin{align}\label{delta_simplified}
%\phi{\delta} = \widetilde{\delta} + \phi \frac{1-K/N}{\lambda}. 
%\end{align}
From the identity in \eqref{delta_j_simplified}, it can be proved that $\widetilde{\delta}$ is the real positive solution, say $\widetilde{\delta}^\star$, of the following third order polynomial equation in \eqref{eq:third_polynomial}.

Since $\frac{\widetilde{\delta}^\star}{\phi} = \frac{1}{N} \tr ({\widetilde{\bf T}})$ and ${\widetilde{\bf T}}$ is diagonal with equal entries, we have $[{\widetilde{\bf T}}]_{k,k}  = \frac{N}{K}\frac{\widetilde{\delta}^\star}{\phi}$ such that \eqref{vartheta_{j}} simplifies to
\begin{align}
    \label{F_{j}_simplified}
    F_{j} &=F = \frac{\kappa}{1+\kappa}\frac{N}{K\phi}\frac{(\tilde{\delta}^\star)^2}{(1+\tilde{\delta}^\star)^2}
\end{align}
and $\Delta_{j}$ in \eqref{delta_{j}} reduces to \eqref{eq:61.delta} where $\vartheta=\vartheta_j=\frac{1}{N}\tr \left({\bf T}_j^2\right)$. To further expand $\vartheta$, we use:
\begin{align}
{\bf T}_j^2\left(\lambda(1+\tilde{\delta}^\star){\bf I}_N+\frac{1}{N}\frac{\overline {\bf{H}}_{jj}\overline {\bf{H}}_{jj}^{H}}{1+\delta^\star \phi}\right)={\bf T}_j
\end{align}
from which it follows that $\lambda(1+\tilde{\delta}^{\star})\vartheta +\frac{1}{N^2} \frac{\tr(\overline {\bf{H}}_{jj}^{H}{\bf T}_j^2\overline {\bf{H}}_{jj})}{1+\delta^\star\phi} =\delta^\star$
or equivalently:
\begin{align}\notag
\vartheta=\frac{\tilde{\delta}^\star}{\phi\lambda(1+\tilde{\delta}^\star)}&+\frac{1-\frac{K}{N}}{\lambda^2(1+\tilde{\delta}^\star)}-\frac{N}{K}\frac{(\tilde{\delta}^\star)^2}{\phi^2}\frac{\kappa}{(1+\kappa)\lambda(1+\tilde{\delta}^\star)^2} \\&-\frac{1}{\lambda^2}\frac{1-\frac{K}{N}}{(1+\tilde{\delta}^\star)^3}\frac{N}{K}\frac{(\tilde{\delta}^\star)^2}{\phi}\frac{\kappa}{1+\kappa}.
\end{align}
By using \eqref{eq:third_polynomial}, we obtain
\begin{align}\notag
\vartheta&=\frac{-\frac{N}{K}\frac{\kappa}{(1+\kappa)\phi}\left(\lambda (\tilde{\delta}^\star)^3+(\tilde{\delta}^\star)^2(\lambda-{\phi}\frac{K}{N}+\phi)\right)}{\phi\lambda^2(1+\tilde{\delta}^\star)^3}\\&+\frac{\tilde{\delta}^\star\phi-\frac{\kappa}{1+\kappa}\tilde{\delta}^\star+\phi}{\phi\lambda^2(1+\tilde{\delta}^\star)^3}.
\end{align}
Putting all these results together, it follows that, if the simplified model in \eqref{simplified_{channel}} is adopted, the SINR asymptotic approximations with RZF and S-MMSE reduce to
\begin{align}\label{overlinegamma_{k}_C}
\frac{1}{ \frac{1}{N\rho^{\rm{dl}}} \frac{1}{\overline\psi}\frac{1}{(1-\lambda\frac{N}{K}\frac{\tilde{\delta}^\star}{\phi})^2} +  \frac{\overline{s}}{\overline\psi(1-\lambda\frac{N}{K}\frac{\tilde{\delta}^\star}{\phi})^2} + {\mathsf{Coherent\; Interf.}}}
\end{align}
where we have used that $\phi_{jlk} = {\frac{\alpha}{1+\kappa}\nu}$ for $l\ne j$. Note that quantities $\overline{s}$ and $\overline{\psi}$ refers to $\overline{s}_{jk}$ and $\overline{\psi}_j$ used in Theorem~\ref{RZF_Theorem}.  
It remains thus to compute $\overline{s}$ and $\overline{\psi}$ in the considered simplified model. 
We begin by treating $\overline{\psi}$. From \eqref{overline{psi}j}, we obtain
\begin{align}\notag
\overline{\psi}  &=\left(\lambda^2\frac{\vartheta}{\phi}\widetilde{\vartheta}{\phi}\frac{N}{K}+\frac{1-F}{\Delta \phi}\frac{N}{K}F\right)^{-1}\!\!\!\!\!\!=\left(-\frac{1}{\phi}\frac{N}{K}+\frac{1-F}{\Delta \phi}\frac{N}{K}\right)^{-1} \\
&=\frac{K\phi}{N}\left(-1+\frac{1}{\Delta}-\frac{\kappa}{1+\kappa}\frac{N}{K\phi}\frac{(\tilde{\delta}^\star)^2}{(1+\tilde{\delta}^\star)^2}\right)^{-1}\!\!\!\notag\\&=\frac{K\nu}{N(1+\kappa)^2}\left(-1+\frac{1}{\Delta}-\frac{\kappa N}{\nu K}\frac{(\tilde{\delta}^\star)^2(1+\kappa)}{(1+\tilde{\delta}^\star)^2}\right)^{-1}\!\!
\end{align}
We now evaluate the interference term. Notice that $\xi_{jk}=\frac{1}{\psi}$ and $\zeta_{jk}=\frac{\lambda^2 N}{K}\frac{\widetilde{\vartheta}}{\phi\Delta}$. Let $d=\frac{1}{1+\kappa}$. 
Hence, after standard calculus we obtain \eqref{eq:C115}
on the top of next page from which \eqref{eq:C:119new} follows.

The computation of the coherent interference requires some preliminaries. From \eqref{eq:third_polynomial}, we obtain\begin{align}
\!\!\!\!\!\!\frac{\phi K}{N}(1+\tilde{\delta}^\star)^2=\lambda\tilde{\delta}^\star(1+\tilde{\delta}^\star)^2+(\tilde{\delta}^\star)^2\phi +\tilde{\delta}^\star \phi +\frac{\tilde{\delta}^\star \kappa}{1+\kappa}
\end{align}
such that
\begin{equation}
\frac{\phi K}{N\lambda}=\tilde{\delta}^\star+\frac{\tilde{\delta}^\star}{\lambda(1+\tilde{\delta}^\star)}\left(\phi+\frac{\frac{\kappa}{1+\kappa}}{\tilde{\delta}^\star+1}\right).
\label{eq:use}
\end{equation}
By using the identity $\frac{\phi K}{N\lambda}=\frac{\phi}{\lambda}+\tilde{\delta}^\star-\phi\delta^\star$ yields $\phi\delta^\star=\frac{\phi}{\lambda(1+\tilde{\delta}^\star)}-\frac{\frac{\kappa}{1+\kappa}\tilde{\delta}^\star}{\lambda(1+\tilde{\delta})^2}$. By using these results, we get
\begin{align}\notag
&\sum_{l\neq j}\phi_{ljk}^2\delta^2\lambda_l^2[\widetilde{\bf T}_l]_{kk}^2\left(1-\lambda_l[\widetilde{\bf T}_j]_{kk}\right)^{-2}=\\&\notag=(L-1)\alpha^2(\delta^\star)^2\phi^2(1+\kappa)^2\left(\frac{1}{\lambda_l[\widetilde{\bf T}_j]_{kk}}-1\right)^{-2}\\
&=(L-1)\alpha^2(1+\kappa)^2\left(\frac{1}{\phi\delta^\star}-\frac{K}{N\lambda \tilde{\delta}^\star\delta^\star}\right)^{-2}\notag\\&=\bigg(\overline{L}-\frac{1}{1+\kappa}\bigg)\frac{\alpha}{\underline{\tau}^2}.
\end{align}

%% file: FINAL_VERSION_arXiv.bbl
\begin{thebibliography}{10}
\providecommand{\url}[1]{#1}
\csname url@samestyle\endcsname
\providecommand{\newblock}{\relax}
\providecommand{\bibinfo}[2]{#2}
\providecommand{\BIBentrySTDinterwordspacing}{\spaceskip=0pt\relax}
\providecommand{\BIBentryALTinterwordstretchfactor}{4}
\providecommand{\BIBentryALTinterwordspacing}{\spaceskip=\fontdimen2\font plus
\BIBentryALTinterwordstretchfactor\fontdimen3\font minus
  \fontdimen4\font\relax}
\providecommand{\BIBforeignlanguage}[2]{{%
\expandafter\ifx\csname l@#1\endcsname\relax
\typeout{** WARNING: IEEEtran.bst: No hyphenation pattern has been}%
\typeout{** loaded for the language `#1'. Using the pattern for}%
\typeout{** the default language instead.}%
\else
\language=\csname l@#1\endcsname
\fi
#2}}
\providecommand{\BIBdecl}{\relax}
\BIBdecl

\bibitem{marzetta2010noncooperative}
T.~L. Marzetta, ``Noncooperative cellular wireless with unlimited numbers of
  base station antennas,'' \emph{IEEE Trans. Wireless Commun.}, vol.~9, no.~11,
  pp. 3590--3600, 2010.

\bibitem{Larsson2014}
E.~Larsson, O.~Edfors, F.~Tufvesson, and T.~Marzetta, ``Massive {MIMO} for next
  generation wireless systems,'' \emph{IEEE Commun. Mag.}, vol.~52, no.~2, pp.
  186 -- 195, February 2014.

\bibitem{Parkvall2017a}
S.~Parkvall, E.~Dahlman, A.~Furuskar, and M.~Frenne, ``Nr: The new {5G} radio
  access technology,'' \emph{IEEE Commun. Std. Mag.}, vol.~1, no.~4, pp.
  24--30, Dec 2017.

\bibitem{Marzetta2016a}
T.~L. Marzetta, E.~G. Larsson, H.~Yang, and H.~Q. Ngo, \emph{Fundamentals of
  {M}assive {MIMO}}.\hskip 1em plus 0.5em minus 0.4em\relax Cambridge
  University Press, 2016.

\bibitem{Bjornson2017abc}
E.~Bj{\"{o}}rnson, J.~Hoydis, and L.~Sanguinetti, ``Massive {MIMO} has
  unlimited capacity,'' \emph{IEEE Trans. Wireless Commun.}, vol.~17, no.~1,
  pp. 574--590, Jan 2018.

\bibitem{massiveMIMOBook}
\BIBentryALTinterwordspacing
E.~Bjornson, J.~Hoydis, and L.~Sanguinetti, ``Massive {MIMO} networks:
  Spectral, energy, and hardware efficiency,'' \emph{Foundations and Trends in
  Signal Processing}, vol.~11, no. 3-4, pp. 154--655, 2017. [Online].
  Available: \url{http://dx.doi.org/10.1561/2000000093}
\BIBentrySTDinterwordspacing

\bibitem{hoydis2013massive}
J.~Hoydis, S.~Ten~Brink, and M.~Debbah, ``Massive {MIMO} in the {UL/DL} of
  cellular networks: How many antennas do we need?'' \emph{IEEE J. Sel. Areas
  Commun.}, vol.~31, no.~2, pp. 160--171, 2013.

\bibitem{Hachem2012}
\BIBentryALTinterwordspacing
W.~Hachem, M.~Kharouf, J.~Najim, and J.~W. Silverstein, ``{A CLT for
  information-theoretic statistics of non-centered Gram random matrices},''
  \emph{{Random Matrices: Theory and Applications}}, vol.~1, no.~2, pp. 1--50,
  2012. [Online]. Available:
  \url{https://hal.archives-ouvertes.fr/hal-00829214}
\BIBentrySTDinterwordspacing

\bibitem{Walid2013}
W.~Hachem, P.~Loubaton, J.~Najim, and P.~Vallet, ``On bilinear forms based on
  the resolvent of large random matrices,'' \emph{Annales de l'Institut Henri
  Poincar, Probabilities et Statistiques}, vol.~49, no.~1, pp. 36 -- 63, 2013.

\bibitem{Bjornson2016z}
E.~Bj{\"{o}}rnson, E.~G. Larsson, and T.~L. Marzetta, ``Massive {MIMO}: ten
  myths and one critical question,'' \emph{IEEE Commun. Mag.}, vol.~54, no.~2,
  pp. 114--123, February 2016.

\bibitem{SanguinettiGC2017}
L.~Sanguinetti, A.~Kammoun, and M.~Debbah, ``Asymptotic analysis of multicell
  massive {MIMO} over {Rician} fading channels,'' in \emph{2017 IEEE
  International Conference on Acoustics, Speech and Signal Processing
  (ICASSP)}, March 2017, pp. 3539--3543.

\bibitem{Wagner12}
S.~Wagner, R.~Couillet, M.~Debbah, and D.~T.~M. Slock, ``Large system analysis
  of linear precoding in correlated {MISO} broadcast channels under limited
  feedback,'' \emph{IEEE Trans. Inf. Theory}, vol.~58, no.~7, pp. 4509--4537,
  July 2012.

\bibitem{Zhang2013a}
J.~Zhang, C.~K. Wen, S.~Jin, X.~Gao, and K.~K. Wong, ``Large system analysis of
  cooperative multi-cell downlink transmission via regularized channel
  inversion with imperfect {CSIT},'' \emph{IEEE Trans. Wireless Commun.},
  vol.~12, no.~10, October 2013.

\bibitem{Sanguinetti_2015c}
L.~Sanguinetti, R.~Couillet, and M.~Debbah, ``Large system analysis of base
  station cooperation for power minimization,'' \emph{IEEE Trans. Wireless
  Commun.}, vol.~15, no.~8, pp. 5480--5496, Aug 2016.

\bibitem{Abla_IT_16}
\BIBentryALTinterwordspacing
H.~Falconet, L.~Sanguinetti, A.~Kammoun, and M.~Debbah, ``Asymptotic analysis
  of downlink {MISO} systems over rician fading channels,'' in
  \emph{Proceedings of the 41st IEEE International Conference on Acoustics,
  Speech and Signal Processing (ICASSP 2016)}, April 2016. [Online]. Available:
  \url{http://arxiv.org/abs/1601.07024}
\BIBentrySTDinterwordspacing

\bibitem{YueL14}
\BIBentryALTinterwordspacing
D.~Yue and G.~Y. Li, ``{LOS}-based conjugate beamforming and power-scaling law
  in {Massive} {MIMO} systems,'' \emph{CoRR}, vol. abs/1404.1654, 2014.
  [Online]. Available: \url{http://arxiv.org/abs/1404.1654}
\BIBentrySTDinterwordspacing

\bibitem{Hoydis11a}
J.~Hoydis, A.~Kammoun, J.~Najim, and M.~Debbah, ``Outage performance of
  cooperative small-cell systems under {Rician} fading channels,'' in
  \emph{IEEE 12th Int. Workshop Signal Processing Advances in Wireless
  Commun.}, June 2011, pp. 551--555.

\bibitem{Zhang13c}
J.~Zhang, C.~K. Wen, S.~Jin, X.~Gao, and K.~K. Wong, ``On capacity of
  large-scale {MIMO} multiple access channels with distributed sets of
  correlated antennas,'' \emph{IEEE J. Sel. Areas in Commun.}, vol.~31, no.~2,
  pp. 133--148, February 2013.

\bibitem{Zhang2014f}
Q.~Zhang, S.~Jin, K.~K. Wong, H.~Zhu, and M.~Matthaiou, ``Power scaling of
  uplink massive {MIMO} systems with arbitrary-rank channel means,'' \emph{IEEE
  J. Sel. Topics Signal Process.}, vol.~8, no.~5, pp. 966--981, Oct 2014.

\bibitem{Said2015}
H.~AL-Salihi and F.~Said, ``Performance evaluation of {Massive MIMO} in
  correlated {Rician} and correlated {Nakagami-m} fading,'' in \emph{2015 9th
  International Conference on Next Generation Mobile Applications, Services and
  Technologies}, Sept 2015, pp. 222--227.

\bibitem{Huang2015a}
Y.~Huang, S.~Ma, and Y.~Wang, ``Uplink achievable rate of full-duplex
  multi-cell {Massive} mimo systems,'' in \emph{2015 IEEE International
  Conference on Communication Workshop (ICCW)}, June 2015, pp. 1131--1136.

\bibitem{Zhang16}
J.~Zhang, L.~Dai, X.~Zhang, E.~Bj{\"{o}}rnson, and Z.~Wang, ``Achievable rate
  of {Rician} large-scale {MIMO} channels with transceiver hardware
  impairments,'' \emph{IEEE Trans. Veh. Technol.}, vol.~65, no.~10, pp.
  8800--8806, Oct 2016.

\bibitem{Masouros2015}
C.~Masouros and M.~Matthaiou, ``Space-constrained {Massive MIMO}: Hitting the
  wall of favorable propagation,'' \emph{IEEE Commun. Lett.}, vol.~19, no.~5,
  May 2015.

\bibitem{Zhang14}
J.~Zhang, C.~Yuen, C.~Wen, S.~Jin, and X.~Gao, ``Ergodic secrecy sum-rate for
  multiuser downlink transmission via regularized channel inversion: Large
  system analysis,'' \emph{IEEE Commun. Lett.}, vol.~18, no.~9, Sept 2014.

\bibitem{Zhang2016a}
J.~Zhang, L.~Dai, S.~Sun, and Z.~Wang, ``On the spectral efficiency of {Massive
  MIMO} systems with low-resolution {ADCs},'' \emph{IEEE Commun. Lett.},
  vol.~20, no.~5, pp. 842--845, May 2016.

\bibitem{Zhang2017a}
J.~Zhang, L.~Dai, Z.~He, S.~Jin, and X.~Li, ``Performance analysis of
  mixed-{ADC} {Massive MIMO} systems over {Rician} fading channels,''
  \emph{IEEE J. Sel. Areas Commun.}, vol.~35, no.~6, pp. 1327--1338, June 2017.

\bibitem{Bjornson2018Rician}
\BIBentryALTinterwordspacing
{\"{O}}.~{\"{O}}zdogan, E.~Bj{\"{o}}rnson, and E.~G. Larsson, ``Massive {MIMO}
  with spatially correlated rician fading channels,'' \emph{CoRR}, vol.
  abs/1805.07972, 2018. [Online]. Available:
  \url{http://arxiv.org/abs/1805.07972}
\BIBentrySTDinterwordspacing

\bibitem{FTufvesson15_measured_MAMIMO}
X.~Gao, O.~Edfors, F.~Rusek, and F.~Tufvesson, ``Massive {MIMO} performance
  evaluation based on measured propagation data,'' \emph{IEEE Trans. Wireless
  Commun.}, vol.~14, no.~7, pp. 3899--3911, July 2015.

\bibitem{Viering2002a}
I.~Viering, H.~Hofstetter, and W.~Utschick, ``Spatial long-term variations in
  urban, rural and indoor environments,'' in \emph{COST273 5th Meeting, Lisbon,
  Portugal}, 2002.

\bibitem{Rusek2013}
F.~Rusek, D.~Persson, B.~K. Lau, E.~Larsson, T.~Marzetta, O.~Edfors, and
  F.~Tufvesson, ``Scaling up {MIMO}: Opportunities and challenges with very
  large arrays,'' \emph{IEEE Signal Process. Mag.}, vol.~30, no.~1, pp. 40 --
  60, Jan 2013.

\bibitem{Kay1993a}
S.~M. Kay, \emph{Fundamentals of Statistical Signal Processing: Estimation
  Theory}.\hskip 1em plus 0.5em minus 0.4em\relax Prentice Hall, 1993.

\bibitem{Ngo2012b}
H.~Ngo, M.~Matthaiou, and E.~G. Larsson, ``Performance analysis of large scale
  {MU}-{MIMO} with optimal linear receivers,'' in \emph{Proc.~IEEE Swe-CTW},
  2012, pp. 59--64.

\bibitem{EmilEURASIP17}
X.~Li, E.~Bj{\"{o}}rnson, E.~G. Larsson, S.~Zhou, and J.~Wang, ``Massive {MIMO}
  with multi-cell {MMSE} processing: Exploiting all pilots for interference
  suppression,'' \emph{EURASIP J. Wireless Commun. Networking}, 2017.

\bibitem{Bjornson2017a}
E.~Bj{\"{o}}rnson, J.~Hoydis, and L.~Sanguinetti, ``Pilot contamination is not
  a fundamental asymptotic limitation in {M}assive {MIMO},'' in
  \emph{Proc.~IEEE ICC}, 2017.

\bibitem{ngo2013energy}
H.~Q. Ngo, E.~G. Larsson, and T.~L. Marzetta, ``Energy and spectral efficiency
  of very large multiuser {MIMO} systems,'' \emph{IEEE Trans. Commun.},
  vol.~61, no.~4, pp. 1436--1449, 2013.

\bibitem{jose2011pilot}
J.~Jose, A.~Ashikhmin, T.~L. Marzetta, and S.~Vishwanath, ``Pilot contamination
  and precoding in multi-cell {TDD} systems,'' \emph{IEEE Trans. Wireless
  Commun.}, vol.~10, no.~8, pp. 2640--2651, 2011.

\bibitem{Kammoun-18}
A.~Kammoun, L.~Sanguinetti, M.~Debbah, and M.-S. Alouini, ``{Asymptotic
  analysis of RZF in large-scale MU-MIMO systems over Rician channels},''
  \emph{submitted to IEEE Trans. Inf. Theory}, 2018.

\bibitem{Ngo2014}
H.~Q. Ngo, E.~G. Larsson, and T.~L. Marzetta, ``Aspects of favorable
  propagation in {Massive} {MIMO},'' in \emph{Proceedings of the 22nd European
  Signal Processing Conference (EUSIPCO)}, Sept 2014, pp. 76--80.

\end{thebibliography}
